\documentclass[a4paper,10pt]{article}
\pdfoutput=1 
             
\usepackage{tikz}
\usetikzlibrary{matrix}
\usepackage{jcappub} 
\usepackage{bm}
\usepackage{hyperref}

\usepackage[T1]{fontenc} 

\usepackage[utf8]{inputenc}
\usepackage{multicol}
\usepackage{lipsum}

\def\d {{d}}

\def\k {\bm{k}}
\def\r {\bm{r}}

\def \apj {\text{ApJ}}

\newcommand{\p}{_{{\text{\tiny$\|$}}}}

\newcommand{\n}{{\bf \hat{n}}}

\newcommand{\HH}{\mathcal{H}}
\newcommand{\two}{^{\text{\tiny \color{green}{({2}})}}}
\newcommand{\one}{^{\text{\tiny \color{red}{({1}})}}}

\newcommand{\<}{\langle}
\renewcommand{\>}{\rangle}

\title{\boldmath Optimal computation of anisotropic galaxy three point correlation function multipoles  using  2DFFTLOG formalism}

\author{Obinna Umeh\note{Corresponding author.}}
\affiliation{Institute of Cosmology \& Gravitation, University of Portsmouth, Portsmouth PO1 3FX, United Kingdom}

\emailAdd{obinna.umeh@port.ac.uk}

\abstract{
We study two key issues militating against the use of the anisotropic three-point correlation function (3PCF)  for cosmological parameter inference: difficulties with its computational estimation and high-dimensionality. We show how high-dimensionality may be reduced significantly by multipole decompositions of all angular dependence. This allows deriving the full expression for the multipole moments of the anisotropic 3PCF and its covariance matrix in a basis where the dimensionality reduces from nine to two at each multipole in the plane-parallel limit.  We use 2D FFTLog formalism to show how the multipole moments with double momentum integrals over the product of bispectrum and two highly oscillating spherical Bessel functions and its covariance with double momentum integrals over the product of three galaxy power spectra and a combination of four highly oscillating spherical Bessel functions may be computed optimally. 

}

\begin{document}
\maketitle
\flushbottom

\section{Introduction}

The three-point correlation function (3PCF) estimates the excess probability of finding three galaxies with locations at the {vertices} of a triangle. It provides an opportunity to probe information that cannot be obtained from the two-point correlation (2PCF), for example;  measurement of the non-linear and tidal bias parameters~\cite{Gil-Marin:2014sta}; probe of the early universe through measurement of various shapes of the primordial non-Gaussianity~\cite{Bartolo:2004if}; probe of the rate of growth of structure under non-linear gravitational
 evolution~\cite{Creminelli:2004pv,Creminelli:2013nua}; 


The most fascinating use of the 3PCF of the large scale structure is as a probe of shapes of non-Gaussinty in the primordial density field predicted by the models of inflation and its alternatives~\cite{Bartolo:2004if,Chen:2006nt}. Models of inflation predict  various shapes of the primordial bispectrum or the inflationary 3PCF~\cite{Scoccimarro:2011pz}. A particular limit of these shapes act as a cosmological collider and could be used to probe the features of high energy particle interaction during inflation on energy scales that can never be achieved on earth~\cite{Arkani-Hamed:2015bza}.   The biggest obstacle in using the galaxy 3PCF to constrain these features is the huge computational overhead associated with computing different permutations of the triangle shapes on large scales and the dependence  of galaxy 3PCF on many variables that make estimating the covariance matrix for cosmological inference a huge task. 

The algorithm that counts triangles is generally not fast, especially on large scales where perturbations theory treatment is possible. In the homogenous and isotropic limit of 3PCF, Slepian \& Eisenstein (2016) \cite{Slepian:2015qza} proposed an algorithm that computes the multipole coefficients of the galaxy 3PCF without explicitly considering triplets of galaxies. The computation time scales like $\mathcal{O}(N^2)$ against $\mathcal{O}(N^3)$ in the traditional approach. This approach builds on the work of~\cite{Szapudi:2004gg}, who first introduced the idea.  The key feature of this formalism involves decomposing the opening angle between any two sides of a triangle in Legendre polynomial 
\begin{eqnarray}
\zeta_g(r_{13}, r_{23},\hat{\r}_{13}\cdot\hat{\r}_{23}) = \sum_{\ell_{13}}\zeta_{g\ell_{13}}(r_{13}, r_{23}) \mathcal{L}_{\ell_{13}}(\hat{\r}_{13}\cdot\hat{\r}_{23})\,,
\label{eq:Szapudi}
\end{eqnarray}
where $\mathcal{L}_{\ell_{13}}$ is the Legendre polynomial, $\zeta_{g\ell_{13}}$ is the multipole moments with respect to the angle between $\hat{\r}_{13}$ and $\hat{\r}_{23}$.  This approach allows immediate insights into the information contained in all triangles by computing only the first few multipoles.  The first detection of the BAO signal using the 3PCF relied on this approach~\cite{Slepian:2016kfz}.  Equation \eqref{eq:Szapudi} is valid in real space, a minimal extension to the monopole of the redshift space bispectrum was done in \cite{Slepian:2016weg}.
The first attempt to extend equation \eqref{eq:Szapudi} to account for anisotropy  was given in~\cite{Slepian:2017lpm,Friesen:2017acf}, where  $\zeta_g$ is decomposed in spherical harmonics basis as
\begin{eqnarray}\label{eq:anisotropic}
\zeta_g({\r}_{13},{\r}_{23}) = \sum_{\ell_1 m_1} \sum_{\ell_2m_2}\zeta_{g\ell_1\ell_2}^{m_1m_2}(r_{13}, r_{23}) Y_{\ell_1 m_1} ({\hat{\r}_{13}})Y^{\star}_{\ell_2 m_2} ({\hat{\r}_{23}})\,,
\end{eqnarray} 
where $Y_{\ell m}$ is the spherical harmonics and $Y^{\star}_{\ell m}$ is the conjugate. The anisotropic 3PCF in equation \eqref{eq:anisotropic} depends on two triangle sides $r_{13}$ and $r_{23}$, the angle each side of the triangle makes with the line of sight. 
Although equation \eqref{eq:anisotropic} provides a complete spherical harmonics basis for decomposing the 3PCF into multipole moments, it does not correspond to the multipole moments of the  redshift space galaxy bispectrum in the well-known Scoccimarro basis~\cite{Scoccimarro:1997st,Scoccimarro:2000sn,Smith:2007sb}.  Recently~\cite{Sugiyama:2018yzo} use a tri-polar spherical harmonics to {decompose} $\zeta_g$ in spherical harmonics
\begin{eqnarray}
\zeta_g({\r}_{13},{\r}_{23},{\n}) 	&=&
	\sum_{\ell_1+\ell_2+L={\rm even}} \zeta_{g\ell_1\ell_2 L}({r}_{13},{r}_{23})\, S_{\ell_1\ell_2L}(\hat{\r}_{13},\hat{\r}_{23},{\n}),
	\label{Eq:TripoSH_J0}
\end{eqnarray}
where $S_{\ell_1\ell_2L}$ is a Tri-polar spherical harmonics~\cite{Varshalovich:book} and the index $L$ is associated with an average over all directions ${\n}$ and in Scoccimarro basis corresponds to the multipole moments of the redshift space bispectrum. The index $\ell_1$ and $\ell_2$ have no obvious physical meaning and there is no guidance on the maximum order of the spherical harmonics to sum in order to recover all the signal.

Our target here is to show  for the first time how  to extend the formalism  introduced in  \cite{Szapudi:2004gg} (equation \eqref{eq:Szapudi}) to anisotropic galaxy 3PCF. Then use the extended formalism to derive  the multipole moments of the anisotropic  galaxy 3PCF in a basis that corresponds to  the Scoccimarro basis for the galaxy  bispectrum in  redshift-space~\cite{Scoccimarro:1997st}. In Scoccimarro basis,  one starts with nine parameters that describe each coordinate of the three triangle vertices, then impose translation invariance, which reduces the nine parameters to six. Imposing the rotation invariance about the line of sight reduces it further to five:  three parameters  characterise the triangle’s shape, e.g. two sides and the enclosed angle, and the remaining two  describe the orientation of the triangle with respect  to the line of sight. It is possible to further reduce the dimensionality to four by averaging over  azimuthal degree of freedom.  This allows to decompose the resulting  anisotropic 3PCF in Legendre polynomials with the  angle between the line of sight and one side of the triangle and the angle between any two sides of the triangle as arguments.


Furthermore,  we show how to optimally compute the double momentum integrals  over the product of the galaxy bispectrum and  two spherical Bessel functions that appear in the expression of the multipole moments of the anisotropic 3PCF using the  2D FFTLog formalism  introduced in~\cite{Fang:2020vhc}\footnote{This is an extension of the 1D FFTLog introduced to cosmology in \cite{Hamilton:1999uv,Hamilton:2015ascl.soft12017H}.}. This formalism allows to  expand the dependence on the wave number in a series of power laws sampled in log-log space. The power laws expansion allows to perform the integral over the spherical Bessel function analytically in terms of the Gamma functions  for the multipoles of the 3PCF and in terms of the the hypergeometric function for its covariance matrix. 

 {We hope that the tools discussed here would be useful in extending the cosmological analysis of Baryon Acoustic Oscillation (BAO)  with data from the extended Baryon Oscillation Spectroscopic Survey (eBOSS) beyond the 2PCF \cite{Wang:2016wjr,Zarrouk:2018vwy,Ross:2016gvb}. Also, the approach we discuss here would be beneficial to the analysis of data  from  future spectroscopic  surveys such as EUCLID \cite{Blanchard:2019oqi}, DESI \cite{Aghamousa:2016zmz}, SKA \cite{Santos:2015gra} etc. }

The rest of the paper is structured as follows: we review the derivation of the multipole moments of 3PCF in real space in section \ref{sec:realspace} and  derive the corresponding expression for the anisotropic 3PCF in sub-section \ref{sec:redshiftspace}. The comparison between our expression for the multipole moments of 3PCF  and previous studies is given  in section \ref{sec:comparism}. We  derive the covariance matrix of the  estimator of the multipole moments of the 3PCF in section \ref{sec:resultsanddis} and conclude in section \ref{sec:conc}.  
We provide details on the implementation of the 2D FFTLog formalism for computing the anisotropic 3PCF in Appendix~\ref{sec:FFTLOG1} and  for the covariance matrix of the multipole moments in Appendix  \ref{sec:FFTLOG2} .  We give  details on  the derivation of the galaxy bispectrum in Appendix \ref{sec:appendix1} and further technical detail on the multipole decomposition is given  in Appendix \ref{sec:detials}.



\noindent
{\bf{Notations}}: 
%
We consider a universe which consists of dark matter and the cosmological constant only, i.e. we ignore the effects of radiation and anisotropic stress tensor.  
{The perturbation theory expansion of any quantity $X$ is normalized as follows}:
$
X = \bar{X} + X\one + X\two/2,
$
where $\bar{X} $ denotes the FLRW  background component.  $X\one$ and $X\two$ are first and second order parts respectively.   We adopt the following values for the cosmological parameters of the standard model~\cite{Ade:2015xua,Aghanim:2018eyx}: Hubble parameter, $h = 0.674$, baryon density parameter, $\Omega_b = 0.0493$,  dark matter density parameter, $\Omega_{\rm{cdm}} = 0.264$,  matter density parameter, 
$\Omega_{m} = \Omega_{\rm{cdm}} + \Omega_b$,  spectral index, $n_s = 0.9608$,  and the amplitude of the primordial perturbation, $A_s = 2.198 \times 10^{9}$.


\section{Galaxy three-point correlation function}\label{sec:derivations}

In this section, we introduced our notations and  the basic ingredients of our approach by reviewing the formalism introduced in \cite{Szapudi:2004gg} for the isotropic 3PCF  before proceeding to the anisotropic  case.

\subsection{Real space galaxy three-point correlation function}\label{sec:realspace}

 The  number of galaxies, $N$, within a given patch of the sky,  ${\rm{d}}\Omega$,   at  a given redshift slice, ${\rm{d}}z$  is given by~\cite{Ellis2009,Challinor:2011bk,Alonso:2015uua}
\begin{eqnarray}\label{eq:numbercount}
 \frac{{\rm{dN}}}{\rm{d} z{\rm{d}}\Omega}  = \bar{n}_g({z})\left[1+ \delta_g({\r})\right] {d}_A^2\frac{\rm{d} \chi}{\rm{d} z},
\end{eqnarray}
where $\chi$ is the comoving distance to the source, $d_A$ is the area distance and  $\bar{n}_g$ is the mean proper  number density of galaxies.  The galaxy density fluctuation $\delta_g$  is related to the matter density fluctuation $\delta_m$ according to the   Eulerian bias model~\cite{Desjacques:2016bnm,Umeh:2019qyd}
\begin{eqnarray}\label{eq:Numbercount2}
\delta_{\rm{g}}({\r}) = b_1\delta_{m}({\r})+ 
\frac{1}{2}\left[b_2\delta_{\rm{m}}({\r}))^2 + b_{\mathcal{K}^2} \mathcal{K}^2({\r})\right]\,,
     \end{eqnarray}
 where $\mathcal{K}^2 = \mathcal{K}_{ij} \mathcal{K}^{ij}$ is the scalar invariant  of the tidal tensor:  $$\mathcal{K}_{ij} ({\r}) = \left(\frac{2}{3\Omega_m\HH}\right)\left[ \partial_i \partial_j \Phi ({\r}) - \frac{1}{3}\nabla^2 \Phi\right] ({\r})\,,$$   $\Phi$  
 is the gravitational potential, it is related to $\delta_{m}$  through the Poisson equation $\delta_{m} =(2/3\Omega_m\HH)  \nabla^2 \Phi$, {where $\HH$ is the conformal Hubble rate}. 
Here, $b_1$, $b_2$ and $ b_{\mathcal{K}^2}$   are the linear, non-linear  and  tidal bias parameters respectively.  
Without loss of generality, we focus on  the clustering bias parameters  for a typical Stage IV spectroscopic survey H$\alpha$ emission line survey~\cite{Yankelevich:2018uaz}
 
\begin{eqnarray}\label{eq:linearbias}
 b_{1} \left( z\right)&=& 0.9 + 0.4 z \, ,
 \\
 b_{2}\left(z\right)&=& -0.704172 -0.207993 z +0.183023 z^{2}-0.00771288 z^3 \, ,\label{eq:Nonlinerbias}
 \\ 
b_{\mathcal{K}^2} \left(z\right)  &=& -\frac{4}{7}(b_{1}(z)-1)
   \label{eq:tidalbias}
\end{eqnarray}
The galaxy 3PCF in configuration space, $\zeta_g$,  is defined as the ensemble average of the galaxy density contrast measured at three {different} points on the sky
 \begin{eqnarray}
\zeta_g({\r}_{1},{\r}_{2},{\r}_{3})&\equiv&\<\delta_{g}({\r}_1)\delta_{g}({\r}_2)\delta_{g}({\r}_3)\> \,,
\\ 
&=& \int \frac{\d^3 k_1}{(2\pi)^3}\int \frac{\d^2 k_2}{(2\pi)^3} 
\int \frac{\d^3 k_3}{(2\pi)^3}
e^{i{\k}_1\cdot {\r}_{1} +i{\k}_2\cdot {\r}_{2} +i{\k}_3\cdot {\r}_{3}}
B_g({\k}_1,{\k}_2,{\k}_3)
 \delta^{D} \left({\k}_1 + {\k}_2 + {\k}_3\right) \,, \qquad 
 \end{eqnarray}
 where we have expanded $\delta_{g}$ in Fourier space and introduced the galaxy bispectrum $B_g$ (for more details on the definition of the galaxy bispectrum see  {equation \eqref{eq:bispectrumgreal} in Appendix \ref{sec:appendix1})}.
The key point to note is that the definition  of $B_g$  includes the  full cyclic permutation of the galaxy density $\delta_g({\k})$ over the three vertices of a triangle:{
\begin{eqnarray}\label{eq:clclicperm}
 B_{g}(  \bm{k}_{1},  \bm{k}_{2},  \bm{k}_{3}) = B^{123}_{g}(  \bm{k}_{1},  \bm{k}_{2},  \bm{k}_{3}) + B^{231}_{g}(  \bm{k}_{2},  \bm{k}_{3},  \bm{k}_{1}) + B^{312}_{g}(  \bm{k}_{3},  \bm{k}_{1},  \bm{k}_{2})\,.
\end{eqnarray}
where the superscript on each $B_{g}$ indicates the corresponding  cyclic permutations of the ${\k}$ indices.}
 The ${\k}'s$ are vectors in Fourier space, its magnitude is related to the wavelength of  mode of the density perturbations, $ \delta^{D} $ is the Dirac delta function which enforces the closure property on the triangle formed by  three Fourier space vectors.
$\zeta_g$ depends on three coordinates (i.e 9 free variables), imposing the translation invariance  reduces to 6 free parameters:
  \begin{eqnarray}\label{eq:pcthreepointfun}
  \zeta_g({\r}_{13},{\r}_{23})  
&=& \int \frac{\d^3 k_1}{(2\pi)^3}\int \frac{\d^3 k_2}{(2\pi)^3} B_g({\k}_1,{\k}_2,-{\k}_1-{\k}_2)e^{i{\k}_1\cdot {\r}_{13} +i{\k}_2\cdot {\r}_{23}}\,.
\label{eq:3pcf}
 \end{eqnarray}
 where  we introduced  a relative distance vector as {${\r}_{MN} = {\r}_M-{\r}_N$},  $M,N$ runs from $1 \cdots 3$.  
The  vectors ${\r}_{MN} $ form sides of a closed triangle  in real space, it satisfies the closure relation: ${\r}_{12}+ {\r}_{23} + {\r}_{31} = 0$. Using the closure property, we  reduce the number of free variables from 6 to 3, i.e two  side lengths and one enclosed angle $r_{13}$, $r_{23}$ and $\hat{\r}_{13}\cdot \hat{\r}_{23}$, where $\nu_{13} =\hat{\r}_{13}\cdot \hat{\r}_{23}$  is  the angle between ${\r}_{13}$ and $ {\r}_{23}$.  In the plane wave {expansion}, we can expand $e^{i{\k}\cdot {\r}_{MN}}$ in Legendre polynomial 
\begin{eqnarray}\label{eq:planewave}
e^{i{\k}\cdot{\r}_{MN}}   = \sum_{\ell }(2\ell+1)i^{\ell}  j_{\ell}(kr_{MN})\mathcal{L}_{\ell}(\hat{\bf{r}}_{MN}\cdot \hat{{\k}})\,,
\end{eqnarray}
where $\mathcal{L}_{\ell}$ is the Legendre polynomial of order $\ell$ and $ j_{\ell}$ is the spherical Bessel function. Again, $B_g$ is the real space galaxy bispectrum, it depends on two amplitudes $k_1$, $k_2$ and the angle between them $\mu_{12} = \hat{\k}_{1}\cdot\hat{\k}_2$\footnote{ Full details on its derivation is given in Appendix \ref{sec:realspacebisptrum}.}.
We expand $\mu_{12}$  and $\nu_{13}$  in Legendre polynomial:
$
B_{g}({k}_1,{k}_2,\mu_{12})= 
\sum_{\ell_{12}}B_{g \ell_{12}}({k}_1,{k}_2)
\mathcal{L}_{\ell_{12}}(\mu_{12}) \,
$
and  $\zeta_g^{123}({r}_{13},{r}_{23},\nu_{13})  = \sum_{\ell_{13}} \zeta_{g  \ell_{13}}({r}_{31},{r}_{23})\mathcal{L}_{\ell_{13}}(\nu_{13})$  respectively.   Using the orthogonality relation  we obtain the real space 3PCF
~\cite{Szapudi:2004gg,Zheng:2004eh,Slepian:2016weg}
\begin{eqnarray}\label{eq:Szapudi5a}
\zeta_{g  \ell_{12}}({r}_{31},{r}_{23})&=& {(-1)^{\ell_{12}} }\int \frac{\d k_1 k^2_1}{2\pi^2}
\int \frac{\d k_2 k^2_2}{2\pi^2}B_{{g}\ell_{12} }(k_1,k_2)
j_{\ell_{12}}(k_1 r_{31})j_{\ell_{12}}(k_2 r_{23})\,.
\end{eqnarray}
At a given $\ell_{12}$, $\zeta_{g  \ell_{12}}$ depends only on  ${r}_{31},$ and ${r}_{23}$.
\begin{figure}[b]
\centering 
\includegraphics[width=155mm,height=100mm] {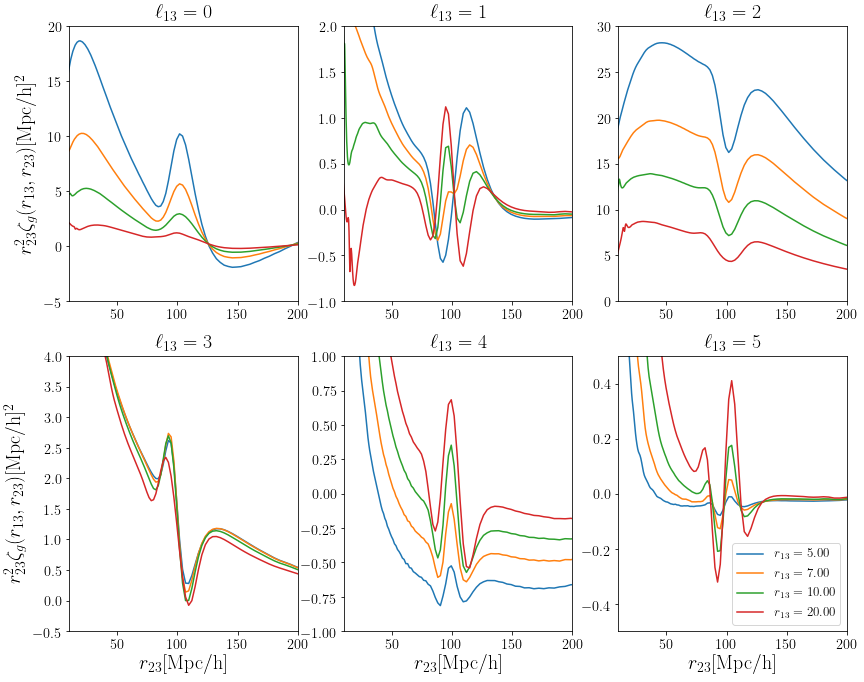}
\caption{We show the first few shape multipoles of the real space  3PCF for the  H$\alpha$ emission line  galaxy. We fixed one side of the triangle to $r_{13} = 5, 7, 10, 20 ~[h^{-1}\rm{Mpc}]=$[blue, yellow, green, red] at $z =1.0$. The monopole  and the quadrupole moments have the highest  amplitude.  The BAO features appear on all the multipoles. }
\label{fig:1Drealspace3pcf}
\end{figure}

%
The numerical computation of equation \eqref{eq:Szapudi5a} has been intractable using traditional methods like the Quadrature.   This is because of  the  double integral over a product of the galaxy bispectrum and highly oscillating product of two spherical Bessel functions.   This constraint motivated earlier works  to focus on the separable  limit where double integrals could be reduced to two independent 1D integrals over a single spherical Bessel function {before the cyclic permutation of the bispectrum given in equation \eqref{eq:clclicperm} is taken}. 
This approach  {which was } initiated in \cite{Scoccimarro:1997st}  {is based} on the realisation that in real space, the galaxy bispectrum is separable  before  cyclic permutation is taken\footnote{The  second order dark matter kernel is separable.}. This leads to the concept of {`pre-cyclic permutation'}  3PCF:
\begin {eqnarray}\label{eq:MonopoleR}
{\zeta}_{g0}^{123}({r}_{31},{r}_{23}) &= & \left[ 2 b_1^2 b_2 + \frac{34}{21} b_1^3\right] \xi_{0}(r_{31})\xi_{0}(r_{23}) \,,
\\
{\zeta}_{g1}^{123}({r}_{31},{r}_{23}) &= &-2b_1^3\left[\xi_{1}^{[-1]}(r_{31})\xi_{1}^{[+1]}(r_{31})
+\xi_{1}^{[-1]}(r_{23})\xi_{1}^{[+1]}(r_{23})\right] \,, \label{eq:DipoleR}
\\
{\zeta}^{123}_{g2}({r}_{31},{r}_{23}) &= &\left[\frac{8}{21} b_1^3 + \frac{2}{3} {b_{\mathcal{K}^2}}b_1^2 \right]\xi_{2}(r_{31})
\xi_{2}(r_{23}) \,,\label{eq:QuadrupoleR}
\end{eqnarray} 
{where ${\zeta}_{g\ell}^{123}$ denotes the 3PCF corresponding to the first  cyclic permutation of the bispectrum defined in equation \eqref{eq:clclicperm}}.  {The double momentum integral} has now reduced to separate 1D integrals which can easily be performed using the traditional quadrature methods
\begin{eqnarray}
\xi_{\ell}(r_{MN})=\int \frac{\d  kk^2}{2\pi^2}{ P_{m}(k) }j_{\ell}(kr_{MN})\,,
\qquad
\xi_{\ell}^{[\pm1]}(r_{MN})=\int \frac{\d k\,  k^2 k^{\pm1}}{2\pi^2} {P_{m} (k)}j_{\ell}(kr_{MN})\,.
\end{eqnarray}
 {The multipole moments of the full 3PCF are  then obtained by taking }the cyclic permutations operations done in real space
\begin{eqnarray}\label{eq:multipole3PCFsep}
{\zeta}_{\ell_{12}} ({r}_{12},{r}_{23}) &=&\frac{(2\ell_{12} +1)}{2} \int_{-1}^{1}\d \nu_{13}
\bigg[\sum_{L_{12}}{\zeta}_{gL_{12}}^{312}({r}_{12},{r}_{23}) \mathcal{L}_{L_{12}} ({\hat{\r}_{12}}\cdot{\hat{\r}_{23}}) 
\\  \nonumber && 
+\sum_{L_{23}}{\zeta}_{gL_{23}}^{231}({r}_{12},{r}_{31}) 
\mathcal{L}_{L_{23}} ({\hat{\r}_{12}}\cdot{\hat{\r}_{31}})
 +\sum_{L_{31}}{\zeta}_{gL_{31}}^{123}({r}_{23},{r}_{31}) \mathcal{L}_{\ell_{31}} ({\hat{\r}_{23}}\cdot{\hat{\r}_{31}})\bigg]
 \mathcal{L}_{\ell_{12}}({\hat{\r}_{13}}\cdot{\hat{\r}_{23}})\,.
\end{eqnarray}
{where  the cosine rule may be used to express  ${\hat{\r}_{12}}\cdot{\hat{\r}_{23}}$  and ${\hat{\r}_{23}}\cdot{\hat{\r}_{31}}$  in terms of $({\hat{\r}_{13}}\cdot{\hat{\r}_{23}})$.}
Extension of this approach to the isotropic limit of the anisotropic 3PCF leads to a complicated results \cite{Slepian:2016weg}.  {This is because of the  angular dependence of the second order redshift space distortion term which involves a $k_3$ wave vector which weakens the separability argument}. 

Our approach avoids this bottleneck by  performing the integration in equation \eqref{eq:Szapudi5a} using 2D FFTLog.  The 2D FFTLog formalism was {introduced} in \cite{Fang:2020vhc}, where it was used to perform the 2D integral that {appears} in the expression for the  non-Gaussian covariance matrix of the {two point correlation function (2PCF)}. We have adapted this for the 3PCF.  We described in detail how this works in Appendix \ref{sec:FFTLOG1}.  Essentially, it involves decomposing  the multipoles of the dimensionless  galaxy bispectrum
${\Delta^{\rm{Real}}_{\rm{B}\ell_{12} }(k_1,k_2) = {k_1^3 k_2^3B_{g\ell_{12} }(k_1,k_2)}/{ (2\pi^2)^2} }$\, in a finite number of power laws sampled in log-log space.  Although the double integrals in equation \eqref{eq:Szapudi5a} run from zero to infinity, use the fact that the integral converges at a finite $k_{\rm{max}}$ to write
\begin{eqnarray}\label{eq:Szapudi6}
\zeta_{g  \ell_{12}}({r}_{13},{r}_{23})&=& i^{2\ell_{12}}\int_{k_{\rm{min}}}^{k_{\rm{max}}}  \frac{\d k_1 }{k_1}
\int_{k_{\rm{min}}}^{k_{\rm{max}}}  \frac{\d k_2 }{k_2} {\Delta^{\rm{Real}}_{{\rm{B}}\ell_{12}}} (k_1,k_2)
j_{\ell_{12}}(k_1 r_{13})j_{\ell_{12}}(k_2 r_{23})\,,
\end{eqnarray}
where we set the $k$-limits to {$[k_{\rm{min}}, k_{\rm{max}}] \propto [10^{-4},2.0]$ [h/Mpc].  } Note that the bispectrum has wiggles  in $k$ due to the Baryon Acoustic oscillation (BAO), therefore,  care must be taken  when choosing $k_{\rm{min}}$ and $k_{\rm{max}}$ such that the k-range  covers scales that contain these features.
{{We show in Figure \ref{fig:1Drealspace3pcf} the first six  multipoles of the 3PCF in real space. We find  strong BAO features in all the multipoles considered. For $\ell_{12} = 0$ and  one side of the triangle kept short, for example  $r_{13}\sim $ short, the amplitude of the monopole becomes negative on large scales. 
Note that the contribution of the shape quadrupole moment is greater than that the shape monopole moment at  say {$r_{13} = 5~[h^{-1}\rm{Mpc}]$} in Figure \ref{fig:1Drealspace3pcf}. This is due to the bias parameters of the H$\alpha$ emission line galaxy we are considering. For this tracer, the nonlinear bias parameter is negative, see equation \eqref{eq:Nonlinerbias}, this reduces the amplitude of the monopole moment as can be seen in equation \eqref{eq:MonopoleR}. The nonlinear bias parameter does not appear  in the expression for the shape  quadrupole moment, see equation \eqref {eq:QuadrupoleR}. The heat map exploring the entire parameter space of $r_{13}$ and $r_{23}$ is given in Figure \ref{fig:heatmap_realspace}.  }}

\begin{figure}[h]
\centering 
\includegraphics[width=155mm,height=100mm] {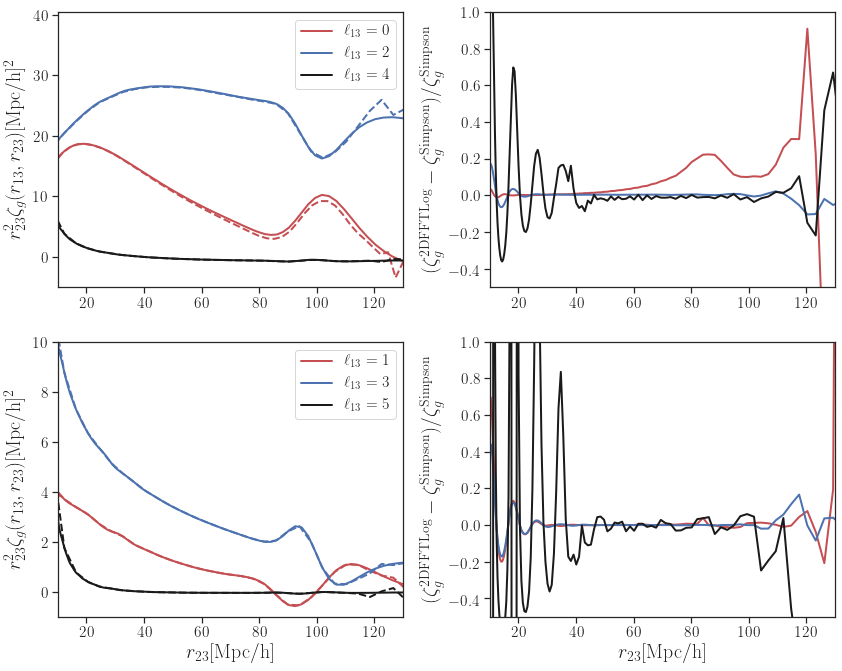}
\caption{{We compare the results of integration of equation \eqref{eq:Szapudi6} using  2D FFTLog to numeral integration using  2D Simpson's rule implementation in python 3.7.  The thicks lines denotes the result from 2D FFTLog implementation while the dash line corresponds to 2D Simpson's rule.
{In both cases, we set $r_{13} = 5 ~[h^{-1}\rm{Mpc}]$ and redshift  at  $z =1.0$.} }}
\label{fig:comparisone3pcf}
\end{figure}

Finally, we compare the performance of the 2D FFTLog integration  to the numerical integration using  2D Simpson's rule implementation of equation \eqref{eq:Szapudi6}.  The comparisons were done with Python 3.7 running on MacBook Pro with 2.3 GHz  Dual-Core Intel Core i5 processor.  We consider just a single shape of the 3PCF with $r_{13} = 5 ~[h^{-1}\rm{Mpc}]$ at  the redshift of one ($z =1.0)$.
It took more than two weeks with 6500 sub-divisions of $[k_{\rm{min}}, k_{\rm{max}}]$ for both $k_1$ and $k_2$  to obtain the results shown in Figure \ref{fig:comparisone3pcf}. In general, there is less than 5\% fractional difference between the two results for separation $20 \le r_{23}\le 120 ~[h^{-1}\rm{Mpc}]$ except for the monopole and $\ell_{13} = 5$. In the case of the monopole, we find that the fractional difference  between the two approaches tends to decrease as we increase the number of sub-divisions but it takes even longer time to complete, hence we stopped at 6500 sub-divisions. 

\subsection{Redshift {space} galaxy three-point correlation function}\label{sec:redshiftspace}

In redshift space, an observer infers the  galaxy position, ${\bf{s}}_{\rm{\tiny{obs}}}$, at a slightly displaced position from the physical position of the galaxy, ${{\r}}$,  due to the effect of the  peculiar velocity according to 
\begin{eqnarray}\label{eq:distortedposition}
{\bm{s}}_{\rm{\tiny{obs}}}& = &{{\r}}   -\frac{1}{\HH}{{\partial_{\p}v} }\, {\n}+ \mathcal{O}(\epsilon)^{2} \,,
\end{eqnarray}
where ${\n}$ is the Line of Sight (LoS) direction, ${{v^i}} = \partial^i v $ is the peculiar velocity of the source  and $\epsilon$ is a small parameter that {controls our perturbative expansion with respect to the background spacetime}. Substituting equation \eqref{eq:distortedposition} in equation \eqref{eq:numbercount}, the number count fluctuations becomes:
\begin{eqnarray}\label{eq:numberfluctuation}
{\Delta_g}(z,{\n})
 &= &  \delta_{g } - \frac{1}{\HH}  \partial_\|^2 v- \frac{2}{\mathcal{H}}\bigg[\delta_{g}\partial_{\parallel}^{2}v + \partial_{\parallel}v\partial_{\parallel}\delta_{g}\bigg] 
 + \frac{2}{\mathcal{H}^{2}}\bigg[\left(\partial_{\parallel}^{2}v\right)^{2} + \partial_{\parallel}v\partial_{\parallel}^{3}v\bigg]\,.
\end{eqnarray}
This is a leading order approximation, the full expression is given in \cite{Bertacca:2014hwa,Yoo:2014sfa,DiDio:2014lka}. We neglected the weak gravitational lensing terms that contribute at the same order as the terms in equation \eqref{eq:numberfluctuation} since we are interested in the Plane-Parallel limit~\cite{Umeh:2015gza,Umeh:2016thy}. {Given  equation \eqref{eq:numberfluctuation}, the 3PCF in redshift space is given by }
\begin{eqnarray}\nonumber 
{\zeta^{\rm{RSD}}_g({\r}_{13},{\r}_{23}, {\n}) }
 &=&\sum_{\ell_1,\ell_2=0}^{\ell_{\text{max}}}{(2\ell_1+1)(2\ell_2+1)} \int \frac{\d k_1 k^2_1}{(2\pi)^3}
\int \frac{\d k_2 k^2_2}{(2\pi)^3}
i^{\ell_1+\ell_2}\int\d {\k}_{{\bot}_{1}}\int \d {\k}_{{\bot}_{2}}
\\ &&
 \times B_{{g}}({\k}_1,{\k}_2,{\n})
j_{\ell_1}(k_1 r_{13})j_{\ell_2}(k_2 r_{23}) \mathcal{L}_{\ell_1}({\hat{\r}}_{13}\cdot{\hat{\k}}_{1})\mathcal{L}_{\ell_2}({\hat{\r}}_{23}\cdot{\hat{\k}}_{2})\,,
\label{eq:3pcf2}
\end{eqnarray}
where we have performed the delta function integral which enforces the closure relation for triangles and {decomposed the angles  ${\r}_{13}\cdot{\k}_1$ and ${\r}_{23}\cdot {\k}_{2}$ in  plane-wave}. 
{$\zeta_g^{\rm{RSD}}({\r}_{13},{\r}_{23}, {\n})$ } depends on six free parameters in plane parallel limit.  Using  the  closure property and rotation with respect to the LoS direction,  we can express $\nu_2 = {\n}\cdot {\r}_{13} $ and $\nu_3 = {\n}\cdot{\r}_{23}$ in terms of $\nu_1$ and $\phi_n$~\cite{Matsubara:1994ApJ}
\begin{eqnarray}
\nu_{2} &=& \nu_{1}\nu_{13} + \sqrt{1 - \nu_{1}^{2} }\sqrt{1-\nu^2_{13}}\cos \phi_{n}\,,
\\
\nu_{3} &=& - \frac{r_{13}}{r_{12}} \nu_{1} - \frac{r_{23}}{r_{12}} \nu_{2} \,,
\end{eqnarray}
where  {$\cos(\nu_{13})  = {\r}_{13}\cdot {\r}_{23}/(r_{13} r_{23})$ is the cosine of the  angle} between ${\r}_{13}$ and $ {\r}_{23}$ and  $\phi_{n}$ is the azimuthal angle that describes the orientation of the triangle with respect to the LoS. This is a configuration space version of the Scoccimarro basis \cite{Scoccimarro:1999ed,Scoccimarro:2000sn}.  At the moment, there are five free parameters that describe $\zeta_g({r}_{13},{r}_{23}, \nu_{13},{\nu}_1,\phi_{n})$.  We can reduce it further to four by averaging over the azimuthal angle leading to the $\phi_n$-average 3PCF or azimuthal  angle average 3PCF
 \begin{eqnarray}
 \zeta_g^{\phi_n}({r}_{13},{r}_{23},\nu_{13},\nu_{1})  =\int_{0}^{2\pi}\frac{\d \phi_n}{2\pi} \zeta_g({r}_{13},{r}_{23},\nu_{13},\nu_{1},{\phi_n}) \,.
 \end{eqnarray}
It is already well-known that the $\phi_n$-averaged galaxy bispectrum are computationally less demanding to estimate given a galaxy catalogue \cite{Bianchi:2015oia,Scoccimarro:2000sn}. This is due to the reduction in dimensionality. 
Also, this limit helps to improve the {signal to noise ratio}~\cite{Bartolo:2004if} and there is a negligible information loss~\cite{Gagrani:2016rfy}.
At this point, we can now expand the $\phi_n$-averaged 3PCF in Legendre polynomial:
$\zeta_g^{\phi_n}({r}_{13},{r}_{23},\nu_{13},\nu_{1})  = \sum_{\ell_{13}} \sum_{L}\zeta_{gL\ell_{13} }^{\phi_n}({r}_{13},{r}_{23}) \mathcal{L}_{\ell_{13}} (\nu_{13})\mathcal{L}_{L }(\nu_{1})\,.$ Similarly, we define the $\phi_{n}$-averaged  galaxy bispectrum 
\begin{eqnarray}
B_{g }^{\phi_n}({k}_1,{k}_2,\mu_{12},\mu_{1}) \equiv \int_{0}^{2\pi} \frac{\d \phi_n}{2\pi} 
B_g({k}_1,{k}_2,\mu_{12},\mu_1,\phi_n)\,
\end{eqnarray}
 and we expand the  enclosed angle in Fourier space  in Legendre polynomial 
$B_{g}^{\phi_n}({k}_1,{k}_2,\mu_{12},\mu_{1}) = \sum_{L = 0}^{\infty}
\sum_{\ell_{12}}B^{\phi_n}_{gL \ell_{12}}({k}_1,{k}_2)\mathcal{L}_{\ell_{12}}(\hat{\k}_1\cdot\hat{\k}_{2}) \mathcal{L}_{L}({\hat{\k}}_{1}\cdot {\n})\,.
$
 The multipoles of $B_{g}^{\phi_n}$ is obtained using the orthogonality condition for the Legendre polynomial
\begin{eqnarray}\label{eq:multipolesofzeta0}
B^{\phi_n}_{gL \ell_{12}}({k}_1,{k}_2) &=&{{(2L+1)\over 2}}\frac{(2\ell_{12} +1)}{2} \int_{-1}^{1}\d\mu_{12}\int_{-1}^{1}\d\mu_{1}\, B_{g}^{\phi_n}({k}_1,{k}_2,\mu_{12},\mu_{1}) 
\\  \nonumber &&
 \times{\cal{L}}_{L } ( \mu_{1} )\mathcal{L}_{\ell_{12}} (\mu_{12})\,.
\end{eqnarray}
Similarly, we obtain the  multipole moments $ \zeta^{\phi_n}_{gL\ell_{13}}$ by decomposing $\mathcal{L}_{L }$ and $\mathcal{L}_{\ell_{13}}$ into spherical harmonics, then using the convolution  theorem of spherical harmonics and the orthonormality relation  we find
 \begin{eqnarray}\label{eq:zetamultipoles}
 \zeta^{\phi_n}_{gL\ell_{13}}({r}_{13},{r}_{23}) 
 &=&\frac{(2L +1)(2\ell_{13} +1)}{(4\pi)^{2}  \mathcal{H}_{\ell_{13},L, \ell_3}}
 \int {d^2\hat{\r}_{13}}\int {d^2\hat{\r}_{23}}\int {d^2\hat{n}} 
 \\ \nonumber &&\times
\sum_{m_{13}Mm_3}
\left( {\begin{array}{ccc}
\ell_{13}& L&  \ell_3\\
  m_{13}& {M}&m_3  \\
 \end{array} } \right)
 Y_{\ell_3 m_3}({\r}_{13})Y^{\ast}_{\ell_{13} m_{13}}({\r}_{23})
Y^{\ast}_{LM}({\n})
\zeta_g^{\phi_n}({r}_{13},{r}_{23}, \nu_{13},\nu_{1})\,,
\end{eqnarray}
where $ Y_{\ell m}$ is the spherical harmonics,  {the bracket is the Wigner} $3j$ symbol, it  satisfies the triangular condition, i.e  {it is zero unless all these conditions are satisfied } $|\ell_{13}-L|\le \ell_{3}\le \ell_{13} +L $  and {$m_{13} + M + m_3= 0$}. We  have introduced $\mathcal{H}_{\ell_{1} \ell_{2} \ell_3}^{0,0,0}$ 
 \begin{eqnarray}
\mathcal{H}_{\ell_{1} \ell_{2} \ell_3}^{0,0,0}&=&\sqrt{\frac{(2\ell_1+1)(2\ell_{2}+1)(2\ell_3+1)}{4\pi}}
\begin{pmatrix}
 \ell_{1}  & \ell_{2} & \ell_{3}\\
  0 & 0 & 0
\end{pmatrix} 
\,.
\end{eqnarray}
Putting  the multipole moment of the galaxy bispectrum in equation \eqref{eq:3pcf2}  and then in  equation \eqref{eq:zetamultipoles} and performing  the following  $\hat{\r}_{13 }$, $\hat{\r}_{23 }$ and ${\n}$  angular integrals  we find
\begin{eqnarray}\label{eq:3pcffinal}
{\zeta}^{\phi_n}_{L\ell_{12}}({r}_{13},{r}_{23}) &=&
i^{\ell_3+\ell_{12}}
\int_{0}^{\infty} \frac{\d k_1 k^2_1}{2\pi^2}
\int_{0}^{\infty}\frac{\d k_2 k^2_2}{2\pi^2}B_{{g}L\ell_{12}  }(k_1,k_2)
 j_{\ell_3}(k_1 r_{13})j_{\ell_{12}}(k_2 r_{23})\,.
\end{eqnarray}
Again ${\zeta}^{\phi_n}_{L\ell_{12}}$ is non-zero only when $|\ell_{12}-L|\le \ell_{3}\le \ell_{12} +L $.  Some of the tools used to simplify the  algebraic steps that lead to equation \eqref{eq:3pcffinal} are given in Appendix \ref{sec:appendix1}.
Choosing the value of $\ell_{3}$ that  saturates the upper bound leads to 
{\begin{eqnarray}\label{eq:zetamoments}
\zeta^{\phi_n}_{gL \ell_{12}}({r}_{13},{r}_{23})&=& i^{2\ell_{12} +L} \int_{0}^{\infty} \frac{\d k_1 }{k_1}
\int_{0}^{\infty} \frac{\d k_2 }{k_2}\Delta^{\rm{RSD}}_{{\rm{B}} L  \ell_{12}}(k_1,k_2)
j_{\ell_{12}+L}(k_1 r_{13})j_{\ell_{12}}(k_2 r_{23})\,,
\end{eqnarray}}
where 
{$\Delta^{\rm{RSD}}_{{\rm{B}}L  \ell_{12}}(k_1,k_2) \equiv {{k_1^3 k_2^3}B^{\phi_n}_{{g} L  \ell_{12}}(k_1,k_2)}/{(2\pi^2)^2 }$} is the dimensionless galaxy bispectrum in redshift space.
We recover exactly the real space 3PCF (equation \eqref{eq:Szapudi5a}) in the isotropic limit  $L=0$ and it agrees with \cite{Szapudi:2004gg}.

\begin{figure}[h]
\centering 
\includegraphics[width=150mm,height=150mm] {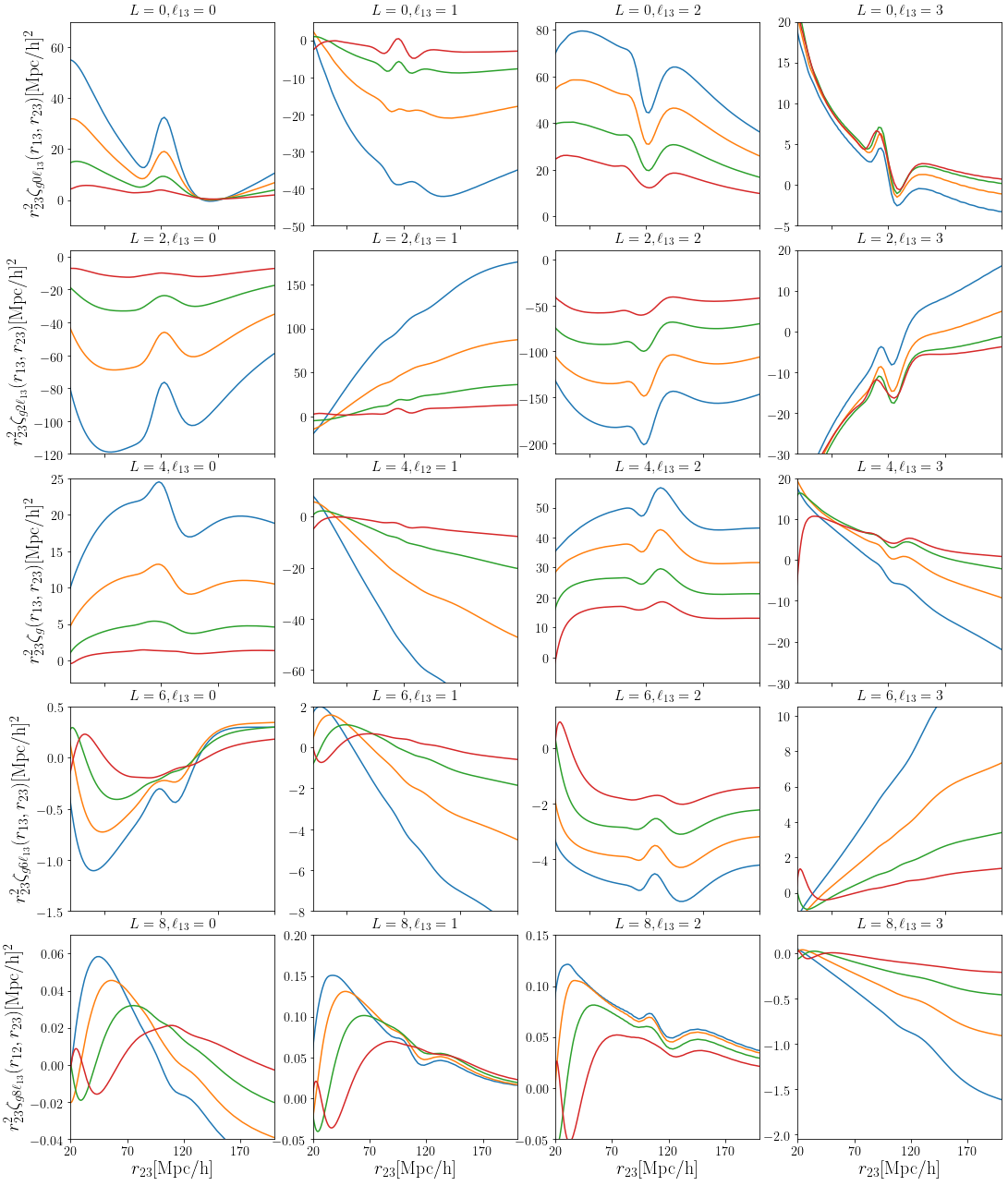}
\caption{\label{fig:redshiftspacemultipoles} The multipoles of the anisotropic 3PCF for the H$\alpha$ emission line galaxy at $z =1.0$ with first few  shape multipoles {$\ell_{13} = 0-3$}.   We set  $r_{13} = [5,7,10,20]=$ [blue, yellow, green, red]. Only even LoS multipoles are induced in the Newtonian limit while  for the shape multipoles both odd and even  are induced.   The BAO features appear in all the  LoS multipoles but it is weaker for  shape odd  multipoles. }
\end{figure}

We describe in detail how equation \eqref{eq:zetamoments} is calculated numerically using the 2D FFTLog formalism  in Appendix  \ref{sec:FFTLOG1}. The multipole moments of the anisotropic 3PCF obtained by integrating equation \eqref{eq:zetamoments}  for the H$\alpha$ emission line galaxy is shown in Figure \ref{fig:redshiftspacemultipoles}.
{We find that  similar to what is obtained in Fourier space in Scoccimarro basis~\cite{Scoccimarro:1997st}, only even LoS  multipoles are induced in the Newtonian limit, while  for the shape multipoles both odd and even multipoles  are induced.   The BAO features appear in all the  LoS multipoles but it is weaker for the first few  shape odd  multipoles. In other words,  the BAO features are more prominent in the first few  shape even multipole moments. The study of the BAO features could focus on those multipoles. {One other important feature to note in Figure \ref{fig:redshiftspacemultipoles} is that the absolute value of the amplitude of the quadrupole moment ($L=2$) is  greater than the  amplitude of the monopole moment ($L=0$). 
This is  likely due to  the nonlinear bias parameter.  The nonlinear  bias parameter for the  H$\alpha$ emission line galaxy  is negative~\cite{Yankelevich:2018uaz}, see also  equation \eqref{eq:Nonlinerbias}.  The negative nonlinear bias parameter  leads to a reduction in the  effective amplitude of the monopole moment relative to the quadrupole moment. This can also be seen analytically  in equation \eqref{eq:MonopoleR}. }


\subsection{Comparison with previous works}\label{sec:comparism}

The most  recent work  on anisotropic 3PCF was given in~\cite{Sugiyama:2018yzo}. The authors  decomposed $\zeta_g$  in tri-polar spherical harmonics as given in equation \eqref{Eq:TripoSH_J0}.
Here we  derive the relationship with the formalism we discussed.  We make use of the  orthonormality condition for $S_{\ell_1\ell_2L}$ 
\begin{eqnarray}\label{eq:tripolarorthonormla}
\int \d^2 \hat{\r}_{13} \int\d^2  \hat{\r}_{23}\int\d^2  {\n}S_{\ell_1\ell_2\ell_3}(\hat{\r}_{13},\hat{\r}_{23},{\n}) 
S_{\ell'_1\ell'_2\ell'_3}(\hat{\r}_{13},\hat{\r}_{23},{\n})  = \delta_{\ell_1\ell'_1}\delta_{\ell_2\ell'_2}\delta_{\ell_3\ell'_3}\,.
\end{eqnarray}
The normalisation of the spherical harmonics used in  \cite{Sugiyama:2018yzo} differs from ours.
The relationship between  $ \zeta_{g\ell_1\ell_2 L}$   and  $\zeta^{\phi_{n}}_{gL \ell_2}$ given in equation \eqref{eq:3pcffinal} maybe obtained by using  equation \eqref{Eq:TripoSH_J0} and \eqref{eq:tripolarorthonormla}
\begin{eqnarray}\label{eq:tripo3pcf}
\zeta_{g\ell_1\ell_2 L}({r}_{13},{r}_{23})
&=&J_{\ell_1 \ell_2 L}
\zeta^{\phi_{n}}_{gL \ell_2}({r}_{13},{r}_{23})\,,
\end{eqnarray}
where $J_{\ell_1 \ell_2 L}$ is a numerical coefficient 
\begin{eqnarray}
J_{\ell_1 \ell_2 L} = \sqrt{\frac{(4\pi)^3(2\ell_1+1)}{(2\ell_{2}+1)(2L+1)}}
\begin{pmatrix}
  \ell_1 & \ell_{2} & L\\
  0 & 0 & 0
  \end{pmatrix}\,.
\end{eqnarray}
Comparing this to equation \eqref{eq:3pcffinal}, we find  that $\ell_2$ corresponds to $\ell_{12}$ and  $\ell_1$ corresponds to $\ell_3$. The 3j symbol is zero except when {$|\ell_{2}-L|\le \ell_{1}\le \ell_{2} +L $} which is in agreement with our result.  Our coefficient differs because \cite{Sugiyama:2018yzo} used a different normalisation for the spherical harmonics. The relationship between equation \eqref{eq:anisotropic} introduced in \cite{Slepian:2017lpm,Friesen:2017acf}  and equation \eqref{eq:tripo3pcf} was discussed in \cite{Sugiyama:2018yzo}. 


\section{Covariance of the multipoles of the  galaxy 3PCF }\label{sec:resultsanddis}

We define the estimator of the azimuthal angled ($\phi_n$-averaged) multipole moments of the radially binned 3PCF as 
\begin{eqnarray}\label{eq:3pcfestimatorappendix}
\hat{\zeta}_{\ell_{13}L}(\bar{r}_{13},\bar{r}_{23}) &\equiv&
\int \frac{d^3{\r}_{13}}{V_{13}}\int \frac{d^3{\r}_{23}}{V_{23}} \int \frac{d \phi_n}{2\pi} {\int \frac{d^2{\n}}{4\pi}}\,
Z_{\ell_{13} L \ell_{3}}(\hat{\r}_{13},\hat{\r}_{23},{\n}) \zeta_{g}({\r}_{13},{\r}_{23}, {\n}) \,,
\end{eqnarray}
 where  $|\ell_{13}-L|\le \ell_{3}\le \ell_{13} +L $,  { the `hat' on thick English alphabets  denotes angular component, we decompose the volume integral into radial and angular components
 \begin{eqnarray}
 \int \frac{d^3{\r}_{13}}{V_{13}} =\frac{1}{V_{13} } \int_{r_{13} -\frac{\Delta r_{13}}{2}}^{r_{!3} +\frac{\Delta r_{13}}{2}} \d r_{13}r^2_{13} \int {\d}^2{\hat{\r}_{13}}\,,		
 \end{eqnarray}}
  and $V_{13} $ is the effective volume of the radial bin
 \begin{eqnarray}
V_{13} = \frac{4\pi}{3} \left( r_{13,\rm{max}}^3 - r^3_{13,\rm{min}}\right) \approx 4\pi r_{13}^2 \Delta r_{13}\,.
 \end{eqnarray} 
We made a thin-bin approximation  in  the second equality.  We introduced the width of the radial bin $\Delta r_{13}$.
Also, we absorbed some of the geometric factors in equation \eqref{eq:3pcfestimatorappendix}  into $Z_{\ell_{13} L \ell_{3}} $   
\begin{eqnarray}
Z_{\ell_{13} L \ell_{3}}(\hat{\r}_{13},\hat{\r}_{23},{\n})  \equiv  \frac{(2L+1)(2\ell_{13} +1)}{\mathcal{H}_{\ell_{13},L, \ell_3} } \sum_{m_{13}Mm_3}
\left( {\begin{array}{ccc}
\ell_{13}& L&  \ell_3\\
  m_{13}& {M}&m_3  \\
 \end{array} } \right)Y^{\ast}_{\ell_3 m_3}(\hat{\r}_{13})Y^{\ast}_{\ell_{13} m_{13}}(\hat{\r}_{23})
Y_{LM}({\n})  \,.
\end{eqnarray}
Furthermore, we define the  anisotropic 3PCF with galaxies at the vertices of the triangle  as 
 \begin{eqnarray}\label{eq:discrete3pcf}
 \zeta_{g}({\r}_{13},{\r}_{23}, {\n}) \equiv \frac{1}{V_s} \int \d^3 r_3
 \Delta_{g}({\r}_{3} + {\r}_{13})\Delta_{g}({\r}_{3} + {\r}_{23})\Delta_{g}({\r}_3)\,,
 \end{eqnarray}
 where $V_s$ is the volume of the survey.   This allows to average  over translations allowing every point in the survey to serve as ${\r}_3$.  There is a clever way of estimating equation \eqref{eq:discrete3pcf} from a  given survey or catalogue  in $\mathcal {O}(N^2)$ time described in \cite{Slepian:2015qza}, further development in this direction was recently reported in \cite{Garcia:2020per}.
From equation \eqref{eq:3pcfestimatorappendix}, we define  the  covariance matrix  of the multipole moments of the azimuthal angle averaged 3PCF as 
 \begin{eqnarray}\label{eq:multipolecovarincedef}
  {{\rm{Cov}}\left[\hat{\zeta}_{\ell_{13}L}(r_{13},r_{23}), \hat{\zeta}_{\ell'_{13}L'}(r'_{13},r'_{23})\right]}&=&
\int\frac{{\d}^2 {r}_{13}}{V_{13} }
\int\frac{{\d}^2{r}_{23}}{V_{23} } \int\frac{{d}^2{r}'_{13}}{V_{13} }
\int\frac{{d}^2 {r}'_{23}}{V_{23} }
\int \frac{d \phi_n}{2\pi} \int \frac{d \phi'_n}{2\pi}\int{ \frac{\d^2 {\n}}{4\pi} }\,\,\,\,
\\&& \times    \nonumber  
\int{\frac{ \d^2 {\n}'}{4\pi} } Z_{\ell_{13} \ell_{3} L} Z_{\ell'_{13} \ell'_{3} L'} {\rm{Cov}}\left[\zeta_g({\r}_{13},{\r}_{23},{\n}) \zeta_g({\r}'_{13},{\r}'_{23},{\n}') \right]\,,
 \end{eqnarray}
In the plane-parallel limit, it is easier to calculate ${\rm{Cov}}\left[\zeta_g({\r}_{13},{\r}_{23},{\n}) \zeta_g({\r}'_{13},{\r}'_{23},{\n}') \right]$ in Fourier space
 \begin{eqnarray}
  {\rm{Cov}}\left[\zeta_g({\r}_{13},{\r}_{23},{\n}) \zeta_g({\r}'_{13},{\r}'_{23},{\n}') \right] 
&=& \int \frac{\d^3 k_1}{(2\pi)^3}\int \frac{\d^3 k_2}{(2\pi)^3} \int \frac{\d^3 k'_1}{(2\pi)^3}\int \frac{\d^3 k'_2}{(2\pi)^3}
 e^{i{\k}_1\cdot {\r}_{13} +i{\k}_2\cdot {\r}_{23}}
 \\  \nonumber && \times
  e^{i{\k}'_1\cdot {\r}'_{13} +i{\k}'_2\cdot {\r}'_{23}} {\rm{Cov}}\left[ B_g({\k}_1,{\k}_2,-{\k}_{12}) B_g({\k}'_1,{\k}'_2,-{\k}'_{12}) \right] \,,
 \end{eqnarray}
where ${\k}_{12} = {\k}_1 +{\k}_2$ and we have made use of the  Fourier transformation {of $\zeta_g$} given in equation \eqref{eq:pcthreepointfun}.
We take the Gaussian limit of  the galaxy bispectrum {covariance given}  in \cite{Sugiyama:2019ike}
  \begin{eqnarray}\nonumber 
{\rm{Cov}}\left[{B}_g({\k}_1,{\k}_2,-{\k}_{12}){B}_g({\k}'_1,{\k}'_2,-{\k}'_{12})\right]
&=&{{\frac{(2\pi)^6}{V_{s}}}} \bigg[\delta^D({\k}_1 +{\k}_1') \delta^D({\k}_2 +{\k}_2') +\delta^D({\k}_2 +{\k}_1') \delta^D({\k}_1 +{\k}_2') 
\\  &&
+ 4{\rm{perms.}}\bigg]
\left(\hat{P}_{g}({\k}_1)\hat{P}_{g}({\k}_2)
\hat{P}_{g}({\k}_{12})\right)\,,
\end{eqnarray}
where $\hat{P}_{g} $ is decomposed into the theory power spectrum and the shot noise: 
$\hat{P}_{g}({\k}) = P_g ({\k})+{1 /{n_g}},$  $P_g$ is the theory galaxy power spectrum and $n_g $ is the galaxy number density.
Since we average over all possible positions of the galaxy at ${\r}_3$ vertex, we focus only on the configuration  that contribute to the covariance matrix  with translation about  the $[{\r}_3,{\r}'_3]$ vertex 
\begin{eqnarray}\label{eq:beforeplaneapprox}
  {\rm{Cov}}\left[\zeta_g({\r}_{13},{\r}_{23},{\n}) \zeta_g({\r}'_{13},{\r}'_{23},{\n}') \right] 
   &=&  \frac{1}{V_s}\int \frac{\d^3 k_1}{(2\pi)^3}\int \frac{\d^3 k_2}{(2\pi)^3}
   \bigg[e^{i k_{1} \left[{\r}_{13} - {\r}'_{13}\right] }
   e^{ik_2 \left[{\r}_{23} - {\r}'_{23}\right]} 
   \\ \nonumber &&
   + e^{i k_{1} \left[{\r}_{13} - {\r}'_{23}\right]} 
   e^{i k_{2} \left[{\r}_{23} - {\r}'_{13}\right]} \bigg]
   \left(\hat{P}_{g}({\k}_1)\hat{P}_{g}({\k}_2)\hat{P}_{g}({\k}_{12})\right)\,.
   \end{eqnarray}
   Expanding the exponentials in plane wave and working through a very lengthy algebra  (see appendix \ref{sec:detials})  gives
   \begin{eqnarray}\label{eq:covariance3pctbody1}
     {{\rm{Cov}}\left[\hat{\zeta}_{\ell_{13}L}(r_{13},r_{23}), \hat{\zeta}_{\ell'_{13}L'}(r'_{13},r'_{23})\right]}
 &=&
    \sum_{L \ell_{12}} 
 \mathcal{O}_{L L'L_2 }^{\ell_{13}\ell'_{13}\ell_{12}  } 
    \int \frac{\d k_1}{k_1} \int \frac{\d k_2}{k_2}
\Delta^{PPP}_{L_2\ell_{12}} ({k_1,k_2} ) 
  \\ \nonumber && \times 
\mathcal{J}_{\ell_{3} \ell'_{3}}^{ \ell_{13} \ell'_{13}}( k_1,k_2, \bar{r}_{13},\bar{r}_{23},\bar{r}'_{13},\bar{r}'_{23})\,,\qquad\quad
\end{eqnarray}
where we have introduced the  multipole moment of the product of  three power spectra
\begin{eqnarray}
\Delta^{PPP}_{L_2\ell_{12}} ({k_1,k_2} ) &\equiv&  \frac{k_1^3k_2^3 \left(\hat{P}({k}_1)\hat{P}({k}_2)\hat{P}({k}_{12})\right)^{\phi_n}_{L_2 \ell_{12}}  }{V_s(2\pi^2)^2} \,.
\end{eqnarray}
The  multipole {moments} of the product of the three power spectra are obtained using
\begin{eqnarray}
 \left(P({k}_1)P({k}_2)P({k}_{12})\right)^{\phi_n}_{L \ell_{12}}   &=& {{(2\ell_{12}+1)\over 2}} {{(2L+1)\over 2}}\int_{-1}^{1} \d \mu_{1} \int_{-1}^{1} \d \mu_{12}  \int_{0}^{2\pi} \frac{{\d}\phi_{n} }{2\pi}
 \\ \nonumber && \qquad    \times ~
  \left(P_{g}({k}_1,\mu_1)P_{g}({k}_2,\mu_2)P_{g}({k}_{12},\mu_3)\right) \mathcal{L}_{L}(\mu_1) \mathcal{L}_{\ell_{12}}(\mu_{12})\,.
\end{eqnarray} 
In the plane-parallel limit for a closed triangle, the three angles are related as {shown} in equation~\eqref{eq:angles}.
In equation \eqref{eq:covariance3pctbody1}, we introduced
\begin{eqnarray}
\mathcal{O}_{L L'L_2 }^{\ell_{13}\ell'_{13}\ell_{12}.   } &\equiv &
    (i)^{\ell_{3} + \ell_{13}}(-i)^{\ell_{3}' +\ell_{13}'}
 \frac{(2L+1)(2\ell_{13} +1)(2L'+1)(2\ell'_{13} +1)}{(4\pi)^2(2L_2+1)(2\ell_{12}+1)}
 \\ \nonumber &&
\times\sum_{L_3=|\ell'_3-\ell_3'|}^{\ell'_3+\ell_3'}
\frac{
\mathcal{H}_{\ell_3 \ell'_{3} L_3}^{0,0,0}
\mathcal{H}_{\ell_{12}L_2  L_3}^{0,0,0}
\mathcal{H}_{\ell_{13} \ell'_{13} \ell_{12}}^{0,0,0}
\mathcal{H}_{L L' L_2}^{0,0,0}}{{ \mathcal{H}^{0,0,0}_{\ell_{13},L, \ell_3}\mathcal{H}^{0,0,0}_{\ell'_{13},L', \ell'_3}}}
\begin{Bmatrix}
   \ell_{13}& L&  \ell_3\\
 \ell'_{13}& L'&  \ell'_3\\
 \ell_{12}&L_2 &L_3
  \end{Bmatrix}\,,
\end{eqnarray}
where the big Curly bracket is the $9j$ symbol~\cite{Varshalovich:book} and  $\mathcal{H}_{\ell_3 \ell'_{3} L_3}^{0,0,0}$ is non-zero only when  $ |L - \ell_{13}| \le \ell_{3} \le L + \ell_{13}$ {and} $| \ell'_{13} - L'| \le \ell'_{3} \le \ell'_{13} + L'$ is satisfied.  We brought together all the spherical Bessel functions under one umbrella in the  thin-shell limit $\Delta r \ll r$:
\begin{eqnarray}\label{eq:SphBessels}
\mathcal{J}_{\ell_{3} \ell'_{3} \ell_{13} \ell'_{13}}( k_1,k_2, \bar{r}_{13},\bar{r}_{23},\bar{r}'_{13},\bar{r}'_{23}) &=&    \bigg[
 j_ {\ell_3}(k_1\bar{r}_{13})
 j_{\ell_3'}(k_1\bar{r}'_{13})  j_ {\ell_{13}}(k_2\bar{r}_{23})
 j_{\ell_{13}'}(k_2\bar{r}'_{23}) 
   \\  && \nonumber 
    + 
      j_ {\ell_3}(k_1\bar{r}_{13}) j_{\ell_3'}(k_1\bar{r}'_{23}) 
   j_ {\ell_{13}}(k_2\bar{r}_{23})j_{\ell_{13}'}(k_2\bar{r}'_{13})
  \bigg] \,.
\end{eqnarray}
%
We show in appendix \ref{sec:FFTLOG2} how to calculate  the  covariance  of the  anisotropic 3PCF multipole moments (equation \eqref{eq:covariance3pctbody1})  using FFTLog formalism.   This formalism  allows to decompose $\Delta^{PPP}_{L_2\ell_{12}} ({k_1,k_2} ) $ into a finite number of power-laws (see equation \eqref {eq:doublepowerlaw}) which then makes it easier to perform the integration over the spherical Bessel functions analytically
\begin{eqnarray}\label{eq:integralDdoubleBessel}
\int \frac{\d k_1}{k_1}{ k_{1}^{\beta_1+i\eta_m}} 
      j_ {\ell_3}(k_1\bar{r}_{13})
 j_{\ell_3'}(k_1\bar{r}'_{13})  =\frac{r_{13}^{-i\eta_{m}}}{r_{13}^{\beta_1}}  \, \int_0^\infty \d x_1\, x_1^{{\beta_1+i\eta_m} -1}j_{\ell_3}(x_1)j_{\ell'_3}(x_1y_1) \,,
\end{eqnarray}
where $x_1 = k_1 \bar{r}_{13}$ and  $ y_1 = \bar{r}'_{13}/\bar{r}_{13}$.  The integral on the RHS  has an analytical solution  in terms of the  hypergeometric function ${}_2F_1$~\cite{Varshalovich:book,Lee:2020ebj}, (see equation \eqref{eq:WSint} for details)
 \begin{eqnarray}
	{I}_{\ell_3 \ell'_3}(\omega_m,y_n) &\equiv &4\pi\int_0^\infty \d x\, x^{\omega_m-1}j_{\ell_3}(x)j_{\ell'_3}(xy_n) \delta_{\ell_{3} \ell'_{3}} 
	\\ 
	&=&\frac{2^{\omega_m-1}\pi^2\Gamma(\ell_3+\frac{\omega_m}{2})}{\Gamma(\frac{3-\omega_m}{2})\Gamma(\ell_3+\frac{3}{2})}\,y_n^{\ell_3}\, 
	{}_2F_1\Bigg[\begin{array}{c} \frac{\omega_m-1}{2}, \ell_3+\frac{\omega_m}{2} , \ell_3+\frac{3}{2}
	\end{array}\Bigg|\, y_n^2\Bigg]\quad (|y_n| \le 1)\, ,\label{eq:WSint}
\end{eqnarray}
where $\Gamma$ is the gamma function and we focus on the limit where $\ell_3 = \ell'_3$. For $|y_n| \ge 1$, we use the following property ${I}_{\ell \ell'}(\omega_m,y_n)  = y_{n}^{\ell} {I}_{\ell \ell'}(\omega_m,1/y_n)$~\cite{Assassi:2017lea}.

\subsection{Estimate of the signal to noise ratio}

Using the 2D FFTLog  formalism (see Appendix \ref{sec:FFTLOG2} for all the technical details on how this is done), we  compute the signal to noise ratio (SNR) for the monopole, quadrupole and hexadecapole of the anisotropic  3PCF for a typical galaxy redshift survey with volume and number density given in Figure  \ref{fig:survey} and the tracer bias parameters are given in {equations \eqref{eq:linearbias} and \eqref{eq:Nonlinerbias}}. 
\begin{figure}[h]
\centering 
\includegraphics[width=100mm,height=80mm] {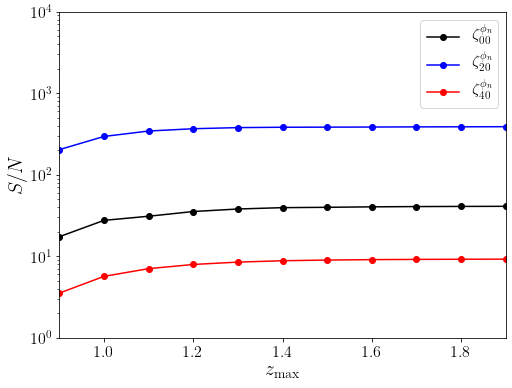}
\caption{\label{fig:SNR}  {This is the  signal to noise ratio for the first three multipole moments of the anisotropic  3PCF, i.e. $L =0,2,4$ with $\ell_{13} = 0$. $\ell_{13}$ corresponds to the monopole limit of the triangular shape.   We set the radial bin width to  $\Delta r_{13} = \Delta r_{23} = 0.5 [{\rm{Mpc}}/h]$ and the range of the radial scales to $[r_{\rm{min}}, r_{\rm{max}}] = [40,120] [{\rm{Mpc}}/h]$.  The  SNR of the  LoS quadrupole moment is much greater than the  LoS monopole. }}
\end{figure}
The survey covers  the redshift range of $0.9$ to $2.0$ and we split it into 10 redshift bins with the width of $0.1$.  
%
We estimate the SNR in the Gaussian limit(see equations \eqref{eq:Gaussianize1} to \eqref{eq:Gaussianize2} on how Gaussian limit is obtained) using 
\begin{eqnarray}
 \left(\frac{S}{N}\right)^2&=&\sum_{z_{\rm{min}}}^{z_{\rm{max}}}\sum_{r_{13} r_{23} }
\frac{{{\zeta}^{\phi_n}_{L\ell_{13}}(\bar{r}_{13},\bar{r}_{23}) }
{{\zeta}^{\phi_n}_{L\ell_{13}}(\bar{r}_{13},\bar{r}_{23}) }}{ {\rm{Var}}\left[\hat{\zeta}_{\ell_{13}L} (\bar{r}_{13},\bar{r}_{23}) \hat{\zeta}_{\ell_{13}L}(\bar{r}_{13},\bar{r}_{23}) \right]}\,.
\end{eqnarray}
{We  neglect the Alcock-Paczynski effects and the loop corrections for simplicity~\cite{Alcock:1979mp}. The cosmological parameters are fixed to the Planck values~\cite{Aghanim:2018eyx} during the computation, the tracer bias parameters evolve according to equations \eqref{eq:linearbias}  and \eqref{eq:Nonlinerbias}}. Without loss of generality, we set $\ell_{13} = 0$ and focus on the first three  even multipole (i.e $L \in $ even). In the rest of the analysis, we set the  radial bin width to {$\Delta r_{13} = \Delta r_{23} = 0.5 $ [Mpc/h]}. The result is shown in Figure \ref{fig:SNR}. 
%
%
Unfortunately, the tree-level approximation of the 3PCF we have used is not  expected to be valid on small scales.  {Therefore, we set {$[r_{13\rm{min}},r_{23\rm{min}}] = [40,40]$ [Mpc/h] and $[r_{13\rm{max}}, r_{23\rm{max}}] = [120,120] [{\rm{Mpc}}/h]$}. Note that the SNR for the LoS  quadrupole moment is greater than the LoS monopole moment. This can easily be understood by looking at the amplitude of the quadrupole moment shown in Figure \ref{fig:redshiftspacemultipoles}. The absolute value of the quadrupole moment of the anisotropic 3PCF is greater than the monopole moment,  this difference get propagated to the SNR as well. }


\section{Conclusion }\label{sec:conc}

We  have shown how the  computational difficulties  associated with the estimation and high-dimensionality of the  anisotropic 3PCF may be alleviated by using the 2D FFTLog formalism and the concept of  decomposing all  its angular dependence  in multipole moments respectively. 
We showed that this can be done in the so-called Scoccimarro basis  which was first introduced in~\cite{Scoccimarro:2000sn} to reduce the  dependence of the anisotropic galaxy bispectrum in Fourier space from nine to five free parameters:  three parameters  characterise the triangle’s shape, e.g. two sides and the enclosed angle, and the remaining two  describe the orientation of the triangle with respect  to the line of sight.
 We showed how to consistently map the anisotropic galaxy bispectrum in this  basis to the corresponding  basis in configuration space for the anisotropic 3PCF,  then we focused exclusively on the limit where the azimuthal degree of freedom is averaged over leading to the azimuthal angle-averaged 3PCF.  Our result is an extension of the formalism introduced in \cite{Szapudi:2004gg} in real space. It was  extended to only the monopole moment (isotropic limit)  in \cite{Slepian:2016weg}. Furthermore, we derived for the first time the  full expression for the covariance  matrix of the azimuthal angled averaged auto 3PCF in plane parallel limit. 

This approach involves expanding  all angular dependence i.e. information associated with triangle shapes  and the information associated with orientation of the triangle  respect to the LoS direction in two different Legendre polynomials respectively.  This allows  for a quick assessment of the total information content of the 3PCF by studying only  the first few multipoles moments. It significantly reduces the need to invest huge resources in evaluating  complicated combinatorial problem associated with different triangle configurations~\cite{Szapudi:2004gg}.  With the  extension  to anisotropic 3PCF that we have shown here,   we are hopeful that this contribution will help to elevate the use of the 3PCF as a reliable tool for cosmological inference  just as the  anisotropic  2PCF especially for probing the  imprint of the primordial non-Gaussainity on the large scale structure. 


Finally, the multipole moments of the galaxy 3PCF we derived contain two momentum  integrals of the product of the galaxy bispectrum and the two spherical Bessel functions. The presence of a double momentum integrals over a  highly oscillating spherical Bessel functions makes it intractable to calculate using the traditional numerical integration algorithm such as Quadrature  or related methods. We have shown how this difficulty maybe circumvented by using  the 2D FFTLog  formalism, which allows  to decompose the dimensionless galaxy bispectrum in a finite number of power-laws  and then perform the integrals over the oscillating spherical Bessel functions analytically. The 2D FFTLog  formalism we use was  developed in \cite{Fang:2020vhc}. It allows to optimally perform these integrals  in $\mathcal{O}\left(N \log N\right)$ times, {where $N$ is
the number of 2D-grid points} .  We show that each multipole moment of the anisotropic 3PCF has a discernible BAO features at the expected scale $100[ \rm{Mpc}/h]$.   In comparison to the 2PCF,  we  find that at each multipole, the amplitude of the  multipole  moment of the 3PCF and its corresponding  BAO bump depends sensitively  on the shape of the triangle. In the limit where one side of the triangle is fixed at small scale, we found a more pronounced BAO peak as we move the fixed side of the triangle deeper into small scales.

\section*{Acknowledgement}
I would like to thank Xiao Fang who developed  2D FFTLog  algorithm, \url{https://github.com/xfangcosmo/2DFFTLog}  that made it possible for us to be able to build on.  Also, I would like to thank Florian Beutler, Rob Crittenden, Daniel Eisenstein, Christain Fidler,  Russell Johnson and Sam Lawrence   for discussions. 
OU  is supported by the 
UK Science \& Technology Facilities Council (STFC) Consolidated Grants Grant ST/S000550/1.


\appendix

\section{Double integration of the 3PCF using  2D FFTLOG }\label{sec:FFTLOG1}

One of the key results of the paper is the full expression for the $\phi_{n}$-averaged anisotropic 3PCF
{\begin{eqnarray}\label{eq:zetamomentsappendix}
\zeta_{gL \ell_{12}}({r}_{12},{r}_{23})&=& i^{2\ell_{12} +L} \int_{0}^{\infty} \frac{\d k_1 }{k_1}
\int_{0}^{\infty} \frac{\d k_2 }{k_2}\Delta^{{\rm{RSD}}}_{{\rm{B}} L  \ell_{12}}(k_1,k_2)
j_{\ell_{12}+L}(k_1 r_{12})j_{\ell_{12}}(k_2 r_{23})\,,
\end{eqnarray}
where have $\Delta^{{\rm{RSD}}}_{{\rm{B}} L  \ell_{12}}$ }is  the dimensionless galaxy bispectrum, $L$ is an integer which indicates the multipole moment with respect to the line of sight, $\ell_{12}$ is an integer which indicates the multipole moment associated with the angle between any two sides of the triangle formed by the galaxy triplet. 

An attempt to perform the integrations in equation \eqref{eq:zetamoments} or  \eqref{eq:Szapudi} using any of the Quadrature methods leads to a sub-optimal result for cosmological inference. This has obvsiouly limited its use despite the amount of cosmological  information  it contains. 
On a close examination, the two biggest stumbling blocks in  performing integration in  equation \eqref{eq:Szapudi} or \eqref{eq:zetamoments} is the presence of the two spherical Bessel functions\footnote{Spherical Bessel function are  roughly  oscillating sine/cosine functions with a decaying amplitude  that scales proportionally to its argument}  in the double integral and the fact that $\Delta^{{\rm{RSD}}}_{{\rm{B}}L  \ell_{12}}$ is not in separable in general.  
Recently, \cite{Fang:2020vhc} proposed a cosmolgically optimal technique to perform the kind of double integrals in equation \eqref{eq:zetamoments}. This technique is based on FFTLOG proposal for 1D integrals proposed in  \cite{TALMAN197835,2009CoPhC.180..332T} and applied to the two-point correlation function in \cite{Hamilton:1999uv}. 
The methods involves  decomposing the dimensionless bispectrum $\Delta^{{\rm{RSD}}}_{{\rm{B}}L  \ell_{12}}(k_1,k_2)$  into a series of products of two power-laws  in the $\log k_1$-$\log k_2$ space~\cite{McEwen:2016fjn,Simonovic:2017mhp,Fang:2019xat,Fang:2020vhc}
\begin{eqnarray}\label{eq:doublepowerlawxi}
  {  \Delta^{{\rm{RSD}}}_{{\rm{B}}L  \ell_{12}}(k_{p},k_{q}) }= \sum_{m=-N_p/2}^{N_p/2}\sum_{n=-N_q/2}^{N_q/2} {c}_{L  \ell_{12} mn}
    \left( \frac{k_{1}^{\beta_1+i\eta_m}}{k_{10}^{i\eta_m}}\right)\left(\frac{k_{2}^{\beta_2+ i\eta_n}}{k_{20}^{i\eta_n} }\right)\,,
    \end{eqnarray}
    where ${c}_{mn}$ is the filtered Fourier coefficients of the dimensional galaxy bispectrum
{\begin{eqnarray}
  {c}_{L  \ell_{12} mn}  = \frac{W_mW_n}{N_pN_q}\sum_{p=0}^{N_p-1}\sum_{q=0}^{N_q-1}\frac{  \Delta^{{\rm{RSD}}}_{{\rm{B}}L  \ell_{12}}(k_{p},k_{q}) }{k_{p}^{\beta_1}k_{q}^{\beta_2}}\exp^{-{2\pi i} \left(\frac{m_p}{N_p}+\frac{n_q}{N_q}\right)}~,
  \end{eqnarray}}
where  $W_m,W_n$ are the $m$-th and $n$-th elements of the 1D window function $\mathbf{W}$ {{introduced to avoid ringing at the edges}}~\cite{McEwen:2016fjn}, $N_x$ is the size of the $k_x$ array,  $\beta_1,\beta_2$ are the real parts of the power law indices for $k_p$ and $k_q$ arrays.    The frequencies  are defined with respect to linear space in $\ln k$
$$\eta_m=2\pi \frac{m}{(N\Delta_{\ln k})} = 2\pi \frac{m}{\log(k_{\rm{max}}/k_{\rm{min}})}\,, \qquad {\rm{with}}\quad \Delta_{\ln k} = \frac{1}{N} \log\left(\frac{k_{\rm{max}}}{k_{\rm{min}}}\right)$$
$\Delta_{\ln k}$ is the linear spacing in $\ln k$.  For example,  given a $k_q$ array, each Fourier mode are separated according to $k_{q} = k_{q0}\exp(q\Delta_{\ln k})$ with $k_{q0}$ being the smallest value in the $k_q$ array.
{The most time consuming part of this calculation is the discretization of {$\Delta^{{\rm{RSD}}}_{{\rm{B}}L  \ell_{12}}(k_1,k_2)$} into an $N_p\times N_q$ grid. There are many options available to help speed up the computation, for example using  a highly optimsed numpy array in python \cite{2020arXiv200610256H} or using  xtensor-stack in c++ with OpenMP support (https://xtensor.readthedocs.io/en/latest/)}.
Using equation \eqref{eq:doublepowerlawxi} in equation \eqref{eq:zetamomentsappendix} gives 
\begin{eqnarray}\label{eq:zetadecomp1}
    \zeta_{gL \ell_{12}}({r}_{12},{r}_{23}) &=& { i^{\ell_{12} +\ell_{3}}}\sum_{m=-N_p/2}^{N_p/2}\sum_{n=-N_q/2}^{N_q/2} \frac{{c}_{L  \ell_{12} mn}} {k_{10}^{i\eta_m}k_{20}^{i\eta_n}}
     \\ \nonumber &&
  \times  \int_0^\infty \frac{dk_1}{k_1}  k_1^{\beta_1+i\eta_m}j_{\ell_{12}}(k_1 r_{13}) \int_0^\infty \frac{dk_2}{k_2}
  k_2^{\beta_2+i\eta_n} 
  j_{\ell_3}(k_2r_{23})\,.
    \end{eqnarray}
    The $k$-integrals may be performed after a change of variables $x_1 = k_1 r_{13}$ and $x_2 = k_2 r_{23}$, for example
    \begin{eqnarray}
     \int_0^\infty \frac{dk_1}{k_1}  k_1^{\beta_1+i\eta_m}j_{\ell_{12}}(k_1 r_{13}) = \frac{r_{13}^{-i\eta_{m}}}{r_{13}^{\beta_1}}  \, \int_{0}^{\infty} {\d x} \,x^{{\beta_1+i\eta_m}-1} j_{\ell}(x) \,.
        \end{eqnarray}
The integral in this form has an analytical solution in terms of the $\Gamma$ functions~\cite{1972hmfw.book.....A}
 \begin{equation}\label{eq:integration1}
    g_{\ell}(\omega_n ) = \frac{4}{\sqrt{\pi}} \int_{0}^{\infty} {\d x} \,x^{\omega_n-1} j_{\ell}(x)= 2^{\omega_n }\frac{\Gamma\left(\frac{\ell+\omega_n }{2}\right)}{\Gamma\left(\frac{3+\ell-\omega_n }{2}\right)}\,.
\end{equation}
For $\omega_m =\beta_1 + i \eta_{m}$ and the range of validity of the bias parameters   are  $-\ell_1<\beta_1<2$ and $-\ell_2<\beta_2<2$.    In order to avoid  singularities in $\Gamma$, non-integer values   for $\beta_1,\beta_2$  are usually chosen. We set $\beta_1 = \beta_2 =1.01$ throughout our computation.  Putting equation \eqref{eq:integration1}  in equation \eqref{eq:zetadecomp1} gives
    \begin{eqnarray}\label{eq:zetadecomp2}
   \zeta_{gL \ell_{13}}({r}_{13},{r}_{23}) 	&=&\frac{\pi  i^{2\ell_{13} +L} }{16 r_{13}^{\beta_1}r_{23}^{\beta_2}}
\sum_{m=-N_p/2}^{N_p/2}\sum_{n=-N_q/2}^{N_q/2}
	 {{c}_{L  \ell_{13} mn}} {(k_{p0}r_{13})^{ -i\eta_m}( k_{q0}r_{23})^{-i\eta_n}} 
	 \\ \nonumber &&
	\times g_{\ell_{13}}(\beta_1+ i\eta_m)g_{\ell_{12} +L}(\beta_2+i\eta_n)~\,.
\end{eqnarray}
Following  \cite{Fang:2020vhc} we take  full advantage of the FFTLog algorithm, we assume that $r_{13},r_{23}$ are identical arrays  and  logarithmically sampled with  a corresponding linear spacing, for example $\Delta_{\ln r_{13}}$ in $\ln r_{13}$, with  $\Delta_{\ln r_{13}}=\Delta_{\ln k_1}$ and $\Delta_{\ln r_{23}}$ in $\ln r_{23}$, with  $\Delta_{\ln r_{23}}=\Delta_{\ln k_2}$, i.e   $r_{13q} = r_{130}\exp(q\Delta_{\ln r})$ with $r_{130}$ being the smallest value in the $r_{13q}$ array. Since the values of $r_{13}$ and $,r_{23}$  are independent of the $\{k_p, k_q\}$ arrays, the summation in equation \eqref{eq:zetadecomp1} can be written as 
    \begin{eqnarray}\label{eq:FFTLOG2DIMP}
   \zeta_{gL \ell_{13}}({r}_{13},{r}_{23}) 	&=&\frac{\pi  i^{2\ell_{12} +L}}{16 r_{p12}^{\beta_1}r_{q23}^{\beta_2}}
   {\rm{IFFT2}}\bigg[
	 {{c}^{\ast}_{L  \ell_{13} mn}} {(k_{p0}r_{120})^{ i\eta_m}( k_{q0}r_{230})^{i\eta_n}} 
	 \\ \nonumber &&
	\times g_{\ell_{12}}(\beta_1- i\eta_m)g_{\ell_{12} +L}(\beta_2-i\eta_n)\bigg]\,.
\end{eqnarray}
 Here $r_{13p},r_{23q}$ ($p,q=0,1,\cdots,N-1$) are the $p$-th and $q$-th elements in the $r_{13}$  and $r_{23}$ array, respectively. We made use of $e^{2\pi i m}$. IFFT2 stands for the two-dimensional Inverse Fast Fourier Transform. We use a modified version of the open source code by \cite{Fang:2020vhc} to calculate equation \eqref{eq:FFTLOG2DIMP}.

\begin{figure}[h]
\centering 
\includegraphics[width=160mm,height=100mm] {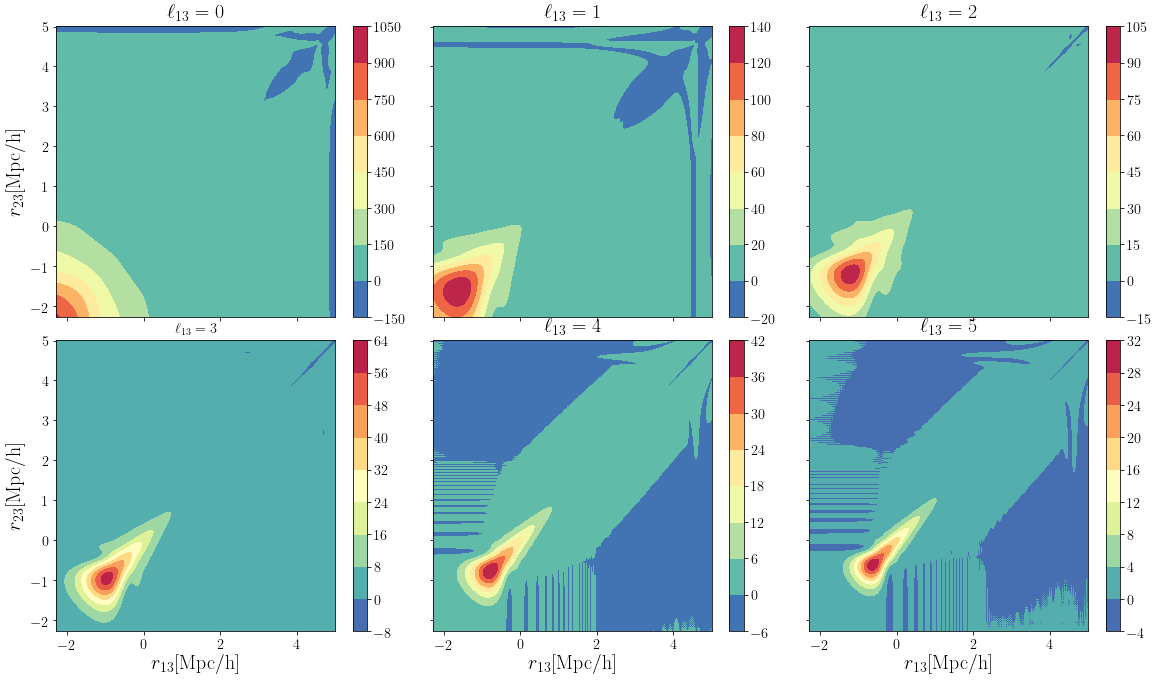}
\caption{\label{fig:i} The 2D heat map in {log-scale} of the  real space 3PCF at different (shape) multipole moment. Here we show the first six multipoles. The features are symmetric about the diagonal in all the multipoles.  This indicates that the 1D plot in Figure \ref{fig:1Drealspace3pcf} captures all the essential features. }
\label{fig:heatmap_realspace}
\end{figure}

\begin{figure}[h]
\centering 
\includegraphics[width=160mm,height=150mm] {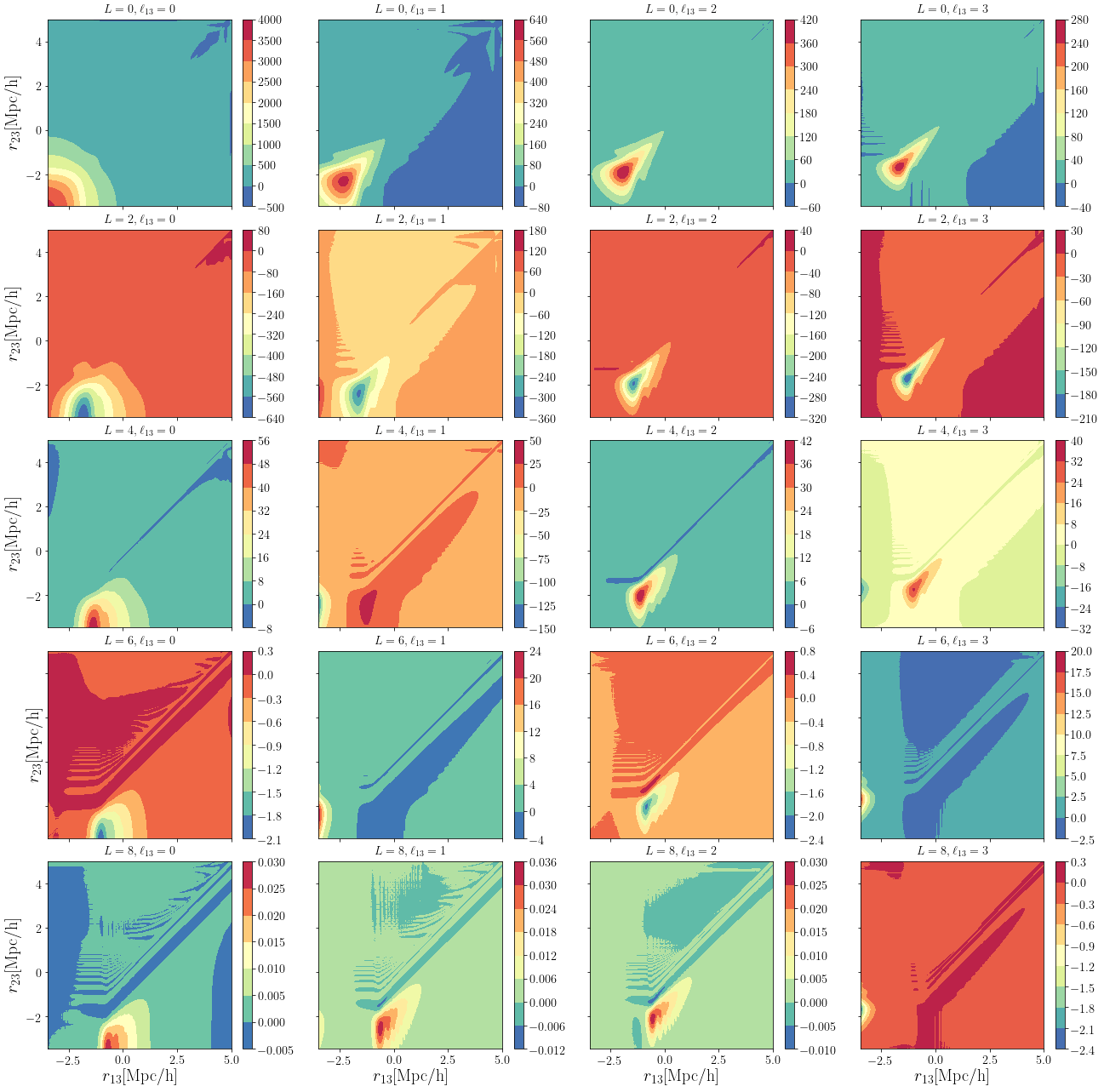}
\caption{\label{fig:i}  The 2D heat map in {log-scale} of the multipoles of the  anisoptropic 3PCF.  In the vertical direction, we have the LOS multipoles starting from  $ L =0$ (top panel) to $L=8$ (bottom panel), while the horizontal direction,  we have the first four multipoles of the triangle configuration.  On the  first left panel we have $\ell_{13} =0$ and last right panel we have $\ell_{13} = 3$. The heat map indicates the amplitude of the 3PCF. Note that the symmetry about the major diagonal  seen in the multipole of the real space 3PCF starts to disappear for the anisotropic 3PCF especially at higher LOS multipole moments. }
\end{figure}
~

\section{Double integration of the  covariance  matrix using 2D FFTLOG }\label{sec:FFTLOG2}

We shall now describe how we calculate the covariance of the multipoles of the 3PCF using 2D FFTLOG formalism. We start from equation \eqref{eq:covariance3pctbody1}
\begin{eqnarray}\nonumber 
 {\rm{Cov}}\left[\hat{\zeta}_{\ell_{13}L}(r_{13},r_{23})\hat{\zeta}_{\ell'_{13}L'}(r'_{13},r'_{23})\right]   &=&
  \sum_{L \ell_{12}} 
 \mathcal{O}_{L L'L_2 }^{\ell_{13}\ell'_{13}\ell_{12}.   } 
    \int \frac{\d k_1}{k_1} \int \frac{\d k_2}{k_2}
\Delta^{PPP}_{L_2\ell_{12}} ({k_1,k_2} ) 
  \\ && \times 
\mathcal{J}_{\ell_{3} \ell'_{3}}^{ \ell_{13} \ell'_{13}}( k_1,k_2, \bar{r}_{13},\bar{r}_{23},\bar{r}'_{13},\bar{r}'_{23})\,,
\label{eq:3pcfcovarianceappendix}
\end{eqnarray}
where $\mathcal{J}_{\ell_{3} \ell'_{3}}^{ \ell_{13} \ell'_{13}}$ is defined in equation \eqref{eq:SphBessels} as a collection of the spherical Bessel functions.
The interest is to evaluate the double k-integrals analytically.  To be able to do this, we perform a power-law decomposition of  $  \Delta^{PPP}_{L_2\ell_{12}}  $
\begin{eqnarray}\label{eq:doublepowerlaw}
  \Delta^{PPP}_{L_2\ell_{12}}   (k_{p},k_{q}) =\sum_{m=-N_p/2}^{N_p/2}\sum_{n=-N_q/2}^{N_q/2} 
  {c}^{PPP}_{L_2  \ell_{12} mn}
    \left( \frac{k_{1}^{\beta_1+i\eta_m}}{k_{10}^{i\eta_m}}\right)\left(\frac{k_{2}^{\beta_2+ i\eta_n}}{k_{20}^{i\eta_n} }\right)\,,
    \end{eqnarray}
    where ${c}^{PPP}_{mn}$ is the filtered Fourier coefficients of the product of three power spectra
\begin{eqnarray}
  {c}^{PPP}_{L_2  \ell_{12} mn}  = \frac{W_mW_n }{N_pN_q} \sum_{p=0}^{N_p-1}\sum_{q=0}^{N_q-1}\frac{ \Delta^{PPP}_{L_2\ell_{12}} (k_{p},k_{q}) }{k_{p}^{\beta_1}k_{q}^{\beta_2}}\exp^{-{2\pi i} \left(\frac{m_p}{N_p}+\frac{n_q}{N_q}\right)}\,,
  \end{eqnarray}
  where $W_m$ is a window function, its proper form is given in \cite{McEwen:2016fjn}.
  Substituting in equation \eqref{eq:3pcfcovarianceappendix} leads to 
    \begin{eqnarray}\label{eq:covariance2}
{\rm{Cov}}\left[\hat{\zeta}_{\ell_{13}L} \hat{\zeta}_{\ell'_{13}L'}\right] 
 &=&
      \sum_{L_2 \ell_{12}} 
 \mathcal{O}_{L L'L_2 }^{\ell_{13}\ell'_{13}\ell_{12}.   } 
\sum_{m=-N_p/2}^{N_p/2}\sum_{n=-N_q/2}^{N_q/2} 
  {c}^{PPP}_{L_2  \ell_{12} mn}
k_{10}^{-i\eta_m} {k_{20}^{-i\eta_n} }
  \\ \nonumber && \times 
     \bigg[    \int \frac{\d k_1}{k_1}{ k_{1}^{\beta_1+i\eta_m}} 
      j_ {\ell_3}(k_1\bar{r}_{13})
 j_{\ell_3'}(k_1\bar{r}'_{13})  \int \frac{\d k_2}{k_2}
     {k_{2}^{\beta_2+ i\eta_n}}
j_ {\ell_{13}}(k_2\bar{r}_{23})
 j_{\ell_{13}'}(k_2\bar{r}'_{23}) 
  \\  \nonumber&& 
 + 
        \int \frac{\d k_1}{k_1}  { k_{1}^{\beta_1+i\eta_m}} j_ {\ell_3}(k_1\bar{r}_{13}) j_{\ell_3'}(k_1\bar{r}'_{23})  \int \frac{\d k_2}{k_2}
      {k_{2}^{\beta_2+ i\eta_n}}
   j_ {\ell_{13}}(k_2\bar{r}_{23})j_{\ell_{13}'}(k_2\bar{r}'_{13})
  \bigg]\,.
\end{eqnarray}
The $k$-integrals over the  spherical Bessel functions in equation \eqref{eq:covariance2} may be calculated analytically after a change of variables; for example $x_1 = k_1 \bar{r}_{13}$, 
\begin{eqnarray}\label{eq:integralDdoubleBessel}
\int \frac{\d k_1}{k_1}{ k_{1}^{\beta_1+i\eta_m}} 
      j_ {\ell_3}(k_1\bar{r}_{13})
 j_{\ell_3'}(k_1\bar{r}'_{13})  =\frac{r_{13}^{-i\eta_{m}}}{r_{13}^{\beta_1}}  \, \int_0^\infty \d x_!\, x_1^{{\beta_1+i\eta_m} -1}j_{\ell_3}(x_1)j_{\ell'_3}(x_1y_n) \,,
\end{eqnarray}
where $ y_1 = \bar{r}'_{13}/\bar{r}_{13}$. Note that there is a freedom to perform the integration via the following change of variable: $x'_1 = k_1 \bar{r}'_{13}$  
\begin{eqnarray}\label{eq:integralDdoubleBessel1}
\int \frac{\d k_1}{k_1}{ k_{1}^{\beta_1+i\eta_m}} 
      j_ {\ell_3}(k_1\bar{r}_{13})
 j_{\ell_3'}(k_1\bar{r}'_{13})  =\frac{{r'}_{13}^{-i\eta_{m}}}{{r'}_{13}^{\beta_1}}  \, \int_0^\infty \d x'_!\, {x'}_1^{{\beta_1+i\eta_m} -1}j_{\ell'_3}(x'_1)j_{\ell_3}(x'_1/y_n) \,.
\end{eqnarray}
This freedom could be understood as a consequence of the Gaussian approximation we have made in the covariance in Fourier space. 
For the integral in the other direction, the following change of variables apply
$ x_{2} = k_2 \bar{r}_{23} $ and  $y_{2} = \bar{r}'_{23}/\bar{r}_{23}$. 
In the integral over the double spherical Bessel functions in equation \eqref{eq:integralDdoubleBessel} has an analytical solution  in terms of the  hypergeometric function ${}_2F_1$~\cite{Varshalovich:book,Lee:2020ebj} 
 \begin{eqnarray}
	{I}_{\ell_3 \ell'_3}(\omega_m,y_n) &\equiv &4\pi\int_0^\infty \d x\, x^{\omega_m-1}j_{\ell_3}(x)j_{\ell'_3}(xy_n) \delta_{\ell_{3} \ell'_{3}} 
	\\ 
	&=&\frac{2^{\omega_m-1}\pi^2\Gamma(\ell_3+\frac{\omega_m}{2})}{\Gamma(\frac{3-\omega_m}{2})\Gamma(\ell_3+\frac{3}{2})}\,y_n^{\ell_3}\, 
	{}_2F_1\Bigg[\begin{array}{c} \frac{\omega_m-1}{2}, \ell_3+\frac{\omega_m}{2} , \ell_3+\frac{3}{2}
	\end{array}\Bigg|\, y_n^2\Bigg]\quad (|y_n| \le 1)\, ,\label{eq:WSint}
\end{eqnarray}
When $y_n >1$,  we use  the property of the hypergeometric function discussed in   \cite{Assassi:2017lea} to compute this region
\begin{eqnarray}
{I}_{\ell \ell'}(\omega_m,y_n)  = y_{n}^{\ell} {I}_{\ell \ell'}(\omega_m,1/y_n) \,,\qquad (|y_n| \ge 1)\,.
\end{eqnarray}
 We use  mpmath implementation of the hypergeometric functions  and the Gamma functions to evaluate ${}_{2}F_1$~\cite{mpmath} and use scipy implementation of the Gamma function to evaluate $\Gamma$  \cite{Scipy:2020NatMe..17..261V}. Also, we verified that result satisfied the recurrence relation  discussed in \cite{Assassi:2017lea}  in  $ y_n < 0.6$ regime $({I}_{\ell \ell'} = {I}_{\ell})$
\begin{eqnarray}
\left(3 + \ell -\frac{\omega_m}{2}\right) I_{\ell +2} (\omega_n , y_n) = \frac{1+ y_n^2}{y_n} \left(\ell+\frac{3}{2}\right) I_{\ell+1} (\omega_m ,y_n) - \left(\ell + \frac{\omega_n}{2}\right) I_{\ell} (\omega, y_n)\,,
\end{eqnarray}
where the initial values are
\begin{eqnarray}
I_{0}(\omega_m,y_n) &=& 2\pi \cos\left(\frac{\pi \omega_m}{2}\right) \Gamma(\omega_m -2) y_n^{-1}\left[\left(1 + y_n\right)^{2-\omega_m}
- \left(1 - y_n\right)^{2-\omega_m} \right]\,,
\\
I_{1}(\omega_m,y_n) & =& \frac{2\pi \cos\left(\frac{\pi \omega_m}{2}\right)\Gamma(\omega_m -2)}{(4-\omega_m) y^2_{n}}
\bigg[\left(1+ y_n\right)^{2-\omega_m} \left( \left(1- y_n\right)^2 + \omega_m y_n \right)
\\ \nonumber &&
-(1-y_n)^{2-\omega_m} \left( (1 +y_n)^2 - \omega_m y_n\right)\bigg]\,.
\end{eqnarray}
 Substituting equation \eqref{eq:WSint} in equation \eqref{eq:covariance2} and  performing some algebraic simplification  leads to 
   \begin{eqnarray}\nonumber 
   {\rm{Cov}}\left[\hat{\zeta}_{\ell_{13}L} \hat{\zeta}_{\ell'_{13}L'}\right] 
 &=&
\frac{1}{{(4\pi)^2}}    \sum_{L \ell_{12}}   \frac{ \mathcal{O}_{L_2 L'L_2 }^{\ell_{13}\ell'_{13}\ell_{12}  } }{{\bar{r}_{13}^{\beta_1} \bar{r}_{23}^{\beta_2}}}
\sum_{m=-N_p/2}^{N_p/2}\sum_{n=-{N_q}/{2}}^{N_q/2}   {c}^{PPP}_{L_2  \ell_{12} mn}
 {(k_{10}\bar{r}_{13})^{-i\eta_m} {(k_{20}\bar{r}_{23})^{-i\eta_n} }}
   \\  && \times 
     \bigg[ 
      { I}_{\ell_3\ell'_3}(\omega_1,y_1) {I}_{\ell_{13} \ell'_{13}}(\omega_2,y_2) 
 + 
     { I}_{\ell_3 \ell'_{3}}(\omega_1,y_3) { I}_{\ell_{13} \ell'_{13}}(\omega_2,y_4) 
  \bigg]\,,
  \label{eq:docmposeCovariance1}
\end{eqnarray}
where $\omega_1 = \beta_1 + i\eta_m$ and $\omega_2 = \beta_2 + i \eta_n $ and we have also defined $ y_3 = \bar{r}'_{23}/\bar{r}_{13}$ and $y_{4} =\bar{ r}'_{13}/\bar{r}_{23}$ to reduce clutter. Again we use  $\bar{r}_{13}$, $\bar{r}_{23}$, $\bar{r}'_{13}$ and $\bar{r}'_{23}$  are arrays logarithmically sampled in linear spacing: $\bar{r}_{13q} = \bar{r}_{130}\exp(q\Delta_{\ln r})$ with $\bar{r}_{130}$ being the smallest value in the $\bar{r}_{13q}$ array. 
Given that in the Gaussian limit, the result is unchanged irrespective of how we perform the radial  integration  i.e  equation \eqref{eq:integralDdoubleBessel} or equation \eqref{eq:integralDdoubleBessel1},  we focus on the limit where $\bar{r}'_{13}=\bar{r}_{13}$ and $\bar{r}'_{23}= \bar{r}_{23}$
 \begin{eqnarray}\label{eq:Gaussianize1}
 y_1 &=& \frac{\bar{r}'_{13}}{\bar{r}_{13}}  =\frac{\bar{r}'_{130}}{\bar{r}_{130}} 
 = y_{10} \,,
 \\
 y_{2} &=& \frac{\bar{r'_{23}}}{\bar{r}_{23}}=\frac{\bar{r}'_{230}}{\bar{r}_{230}} 
 = y_{20} \,,\\
  y_3 &=&\frac{ \bar{r}'_{23}}{\bar{r}_{13}}=\frac{\bar{r}'_{230}}{\bar{r}_{130}} e^{\left( m-n\right)\Delta \ln r}  
 = y_{30}e^{\left( m-n\right)\Delta \ln r}  = \hat{y}_{30}\,, \qquad {\rm{with}} \qquad y_{30} = \frac{\bar{r}'_{230}}{\bar{r}_{130}} = \frac{\bar{r}_{230}}{\bar{r}_{130}} \,,
  \\
  y_{4} &=&\frac{\bar{ r}'_{13}}{\bar{r}_{23}}=\frac{\bar{r}'_{130}}{\bar{r}_{230}} e^{\left( m-n\right)\Delta \ln r}  
 = y_{40}e^{\left( m-n\right)\Delta \ln r}  = \hat{y}_{40}\,, \qquad {\rm{with}} \qquad y_{40} = \frac{\bar{r}'_{130}}{\bar{r}_{230}} = \frac{\bar{r}_{130}}{\bar{r}_{230}} \,.
 \label{eq:Gaussianize2}
 \end{eqnarray}
%
Implementing these in equation \eqref{eq:docmposeCovariance1} leads to 
   \begin{eqnarray}\nonumber 
   {\rm{Cov}}\left[\hat{\zeta}_{\ell_{13}L} \hat{\zeta}_{\ell'_{13}L'}\right] 
 &=&
\frac{1}{{(4\pi)^2}}    \sum_{L_2 \ell_{12}}   \frac{ \mathcal{O}_{L L'L_2 }^{\ell_{13}\ell'_{13}\ell_{12}.   } }{{r_{13}^{\beta_1} r_{23}^{\beta_2}}}
\sum_{m=-N_p/2}^{N_p/2}\sum_{n=-{N_q}/{2}}^{N_q/2}   {c}^{PPP}_{L_2  \ell_{12} mn}
 {(k_{10}r_{130})^{-i\eta_m} {(k_{20}r_{230})^{-i\eta_n} }}
   \\  && \times 
     \bigg[ 
      { I}_{\ell_3\ell'_3}(\omega_1,{y}_{10}) {I}_{\ell_{13} \ell'_{13}}(\omega_2,{y}_{20}) 
 + 
     { I}_{\ell_3 \ell'_{3}}(\omega_1,\hat{y}_{30}) { I}_{\ell_{13} \ell'_{13}}(\omega_2,\hat{y}_{40}) 
  \bigg]\,,
\end{eqnarray}
We sum over the $N_p\times N_q$ samples using inverse FFT in $N\log N$ times
 \begin{eqnarray}\nonumber 
   {\rm{Cov}}\left[\hat{\zeta}_{\ell_{13}L} \hat{\zeta}_{\ell'_{13}L'}\right] 
 &=&
 \frac{1}{{(4\pi)^2}}       \frac{1 }{{r_{13}^{\beta_1} r_{23}^{\beta_2}}}
    {\rm{IFFT2}}\bigg\{\mathcal{C}^{\ell_{13} \ell'_{13} L L' }_{mn}
 {(k_{10}r_{130})^{i\eta_m} {(k_{20}r_{230})^{i\eta_n} }}
 \qquad \qquad
   \\  && \times 
\bigg[   { I}_{\ell_3\ell'_3}(\beta_1 - i\eta_m, r'_{130}/r_{130})  {I}_{\ell_{13} \ell'_{13}}(\beta_2 - i \eta_n, r'_{230}/r_{230}) 
\\ \nonumber &&
 +     
     { I}_{\ell_3 \ell'_{3}}(\beta_1 - i\eta_m,r'_{23}/r_{13}) { I}_{\ell_{13} \ell'_{13}}(\beta_2 - i \eta_n, r'_{13}/r_{23}) 
  \bigg]\bigg\}\,,
\end{eqnarray}
The most computationally demanding part is $\mathcal{C}^{\ell_{13} \ell'_{13} L L' }_{mn}$, since we  need to sum over all $L_{2}$ and the first few multipoles of $\ell_{12}$
\begin{eqnarray}\label{eq:multipolesmoments}
\mathcal{C}^{\ell_{13} \ell'_{13} L L' }_{mn}& =& \sum_{L_2 \ell_{12}}^{8,3} \mathcal{O}_{L L'L_2 }^{\ell_{13}\ell'_{13}\ell_{12}.   }
  {c}^{PPP\star }_{L_2  \ell_{12} mn}
  \\
  &=& \mathcal{O}_{L L' 0 }^{\ell_{13}\ell'_{13}0   } {c}^{PPP\star }_{00 mn} + \mathcal{O}_{L L'0 }^{\ell_{13}\ell'_{13}1  } {c}^{PPP\star }_{01 mn}
  + \mathcal{O}_{L L'0 }^{\ell_{13}\ell'_{13}2   } {c}^{PPP\star }_{02 mn}+  \mathcal{O}_{L L'2}^{\ell_{13}\ell'_{13} 0   } {c}^{PPP\star }_{20 mn} 
  \\  \nonumber &&
  + \mathcal{O}_{L L' 2 }^{\ell_{13}\ell'_{13}1} {c}^{PPP\star }_{21 mn} 
  + \mathcal{O}_{L L' 2 }^{\ell_{13}\ell'_{13}2   }  {c}^{PPP\star }_{22mn} + 
  \mathcal{O}_{L L'4 }^{\ell_{13}\ell'_{13}0  } {c}^{PPP\star }_{40mn}+ 
  \mathcal{O}_{L L'4 }^{\ell_{13}\ell'_{13}1  } {c}^{PPP\star }_{41 mn} 
    \\  \nonumber &&
  +  
  \mathcal{O}_{L L'4}^{\ell_{13}\ell'_{13} 2  } {c}^{PPP\star }_{42 mn} + \cdots
\end{eqnarray}
Note that $  {c}^{PPP\star }_{L_2  \ell_{12} mn} = 0$ for $L_2 \in {\rm{odd}}$. This part of the computation could be improved further by using OpenMP or GPU. This is beyond the scope of the current project but it is an interesting direction to pursue. 
We made use of  the implementation of the SymPy/Wigner package~\cite{Sympy:10.7717/peerj-cs.103} to compute 3j and  9j symbols.
%

\begin{figure}[th]
\centering 
\includegraphics[width=80mm,height=60mm] {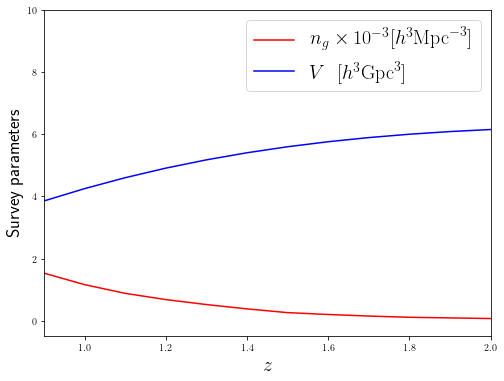}
\caption{\label{fig:survey} We show the mean galaxy number density and the survey volume.  The survey covers about $0.38$ fraction of the sky and the redshift range of $0.9$ to $2.0$. This is the information used in estimating the SNR.}
\end{figure}

\section{Galaxy bispectrum: shape and anisotropic  multipoles}\label{sec:appendix1}

\subsection{Multipoles of the real space galaxy bispectrum}\label{sec:realspacebisptrum}

The galaxy density fluctuation $\delta_g$  is related to the matter density fluctuation $\delta_m$ according to the   Eulerian bias model ~\cite{Desjacques:2016bnm,Umeh:2019qyd}
\begin{eqnarray}\label{eq:Numbercount2}
\delta_{\rm{g}}({\r}) = b_1\delta_{m}({\r})+ 
\frac{1}{2}\left[b_2\delta_{\rm{m}}({\r}))^2 + b_{\mathcal{K}^2} \mathcal{K}^2({\r})\right]\,,
     \end{eqnarray}
 where $b_1$, $b_2$ and $ b_{\mathcal{K}^2}$   are the linear, non-linear  and  tidal bias parameters respectively.  $\mathcal{K}^2 = \mathcal{K}_{ij} \mathcal{K}^{ij}$ is a scalar invariant tidal field constructed from the  tidal tensor,  At second order, we use the expression for the dark  matter density field during the matter dominance
 \begin{eqnarray}\label{eq:matterdensity}
\delta\two({\r}) &=& \frac{34}{21} (\delta\one)^2_m({\r}) - 2{\partial^j\nabla^{-2} \delta\one_{\rm{m}}}({\r}) \partial_j \delta\one_m({\r}) + \frac{4}{7}\mathcal{K}^2 ({\r}) \,.
\end{eqnarray}
During the  $\Lambda$CDM era, there are redshift dependent terms that appear in the coefficient of the terms above, however, equation \eqref{eq:matterdensity} remains a good approximation~\cite{Villa:2015ppa}.
 In Fourier space $\delta_{g}({\k}) = \delta_{g}\one({\k})+ \delta_{g}\two({\k})$, with $\delta_{g}\one({\k}) = b_1 \delta_{m}\one({\k})$ and 
 \begin{eqnarray}
\delta_{g}\two({\k}) &=&
\int \frac{\d^3 k_1}{(2\pi)^3}\frac{\d^3 k_2}{(2\pi)^3}\delta_m({\k}_1) \delta_m({\k}_2)  \mathcal{K}_{\rm{R}}^{(2)}({\k}_1,{\k}_2)
(2\pi)^3\delta^D\left({\k}_1 + {\k}_2 - {\k}\right)\,,
\end{eqnarray}
where the second order momentum space kernel is given by
\begin{eqnarray}
\mathcal{K}_{\rm{R}}^{(2)}(\bm{k}_{1}, \bm{k}_{2})
&=& b_{2}+ b_{1}F_{2}({\k}_{1},{\k}_{2})+ 
b_{\mathcal{K}^2}\mathcal{K}_{2}({\k}_{1},{\k}_{2})\,,
\end{eqnarray}
where  $ F_2$ and $\mathcal{K}_2$ are the Fourier space kernel for the dark matter density and tidal field
\begin{eqnarray}
 F_2({\k}_1,{\k}_2)&=&\frac{5}{7}+\frac{1}{2}\frac{{\k}_1 \cdot {\k}_2}{k_1
k_2}\left(\frac{k_1}{k_2}+\frac{k_2}{k_1}\right)+\frac{2}{7}
\left(\frac{{\k}_1 \cdot {\k}_2}{k_1 k_2}\right)^2\,,
\\
\mathcal{K}_2({\k}_1,{\k}_2)& =& \frac{\left({
 \k}_1\cdot {\k}_2\right)^2}{\left(k_1 k_2\right)^2} - \frac{1}{3}\,.
\end{eqnarray}
In perturbation theory, the tree level galaxy bispectrum may be calculated from two first order  galaxy density constrast
 and one second order galaxy density contrast
 \begin{eqnarray}\label{eq:bispectrumwithcyl}
\big\langle \delta_{g}( \bm{k}_{1}) \delta_{g}( \bm{k}_{2}) \delta_{g}( \bm{k}_{3})\big \rangle &=&\frac{1}{2} \big\langle \delta_{g}\one( \bm{k}_{1}) 
\delta_{g}\one( \bm{k}_{2}) \delta_{g}\two( \bm{k}_{3})\big \rangle +   \text{2 cy. p.} 
\end{eqnarray}
where we have included 2 clyic perturbation since the second order galaxy density contrast can occupy the first two slots as well.  
The real space galaxy bispectrum is given by
\begin{eqnarray}
B_{g}( {\k}_{1},  {\k}_{2},  {\k}_{3}) 
= \mathcal{K}\two_{R}({\k}_{1},  {\k}_{2}) P_{m}(k_{1})P_{m}(k_{2}) +   \text{2 cy. p.} 
\label{eq:bispectrumgreal}
\end{eqnarray}
where $ P_{m}(k)$ is the matter power spectrum. We can expand the  angular dependence  of $B_g$  into Legendre polynomial  
\begin{eqnarray}\label{eq:Bgmultipoles}
B_{g}({k}_1,{k}_2,\mu_{12}) &=&
\sum_{\ell_{12}}B_{g \ell_{12}}({k}_1,{k}_2)
\mathcal{L}_{\ell_{12}}(\mu_{12}) \,, 
\end{eqnarray}
  Using the orthogonality condition we obtain the multipoles of the galaxy bispectrum 
\begin{eqnarray}\label{eq:Bmultipoles1}
B_{g  \ell_{12}}({k}_1,{k}_2) &=& {{(2\ell_{12}+1)\over 2}} \int_{-1}^{1}\d\mu_{12}
B_g({k}_1,{k}_2,\mu_{12}) \mathcal{L}_{\ell_{12}}(\mu_{12}) \,.
\end{eqnarray}

\subsection{Shape multipoles of  the anisotropic galaxy bispectrum}\label{sec:redshiftspacebisptrum}

Expanding equation \eqref{eq:numberfluctuation} in  Fourier space and in plane-parallel limit gives
\begin{eqnarray}\label{eq:galaxyoverfnl}
\Delta_{\rm{N} }({\k})&=&{\mathcal{K}}\one_{\rm{g} } ({\k}) \delta_{m}({\k}) 
\\ \nonumber &&
+ \frac{1}{2}\int \frac{d^3k_1}{(2\pi)^3}\frac{d^3k_2}{(2\pi)^3}{\mathcal{K}_{\rm{g}}}\two ({\k}_{1}, {\k}_{2},{\k}_3)\delta_{m}({\k}_1)\delta_{m}({\k}_2)(2\pi)^3 \delta^{(3)}\left({\k} -{\k}_1 -{\k}_2\right)
 \,.
\end{eqnarray}
where $\mathcal{K}_{\rm{N}}$ is a Fourier space kernel for the galaxy density in redshift space, we have separated $\mathcal{K}_{\rm{N}}$  into linear and the second order part gives
\begin{eqnarray}
\mathcal{K}_{\rm{N}}\one(k_{1})& =& b_1+f\mu^2_{1}\,,
\\
\mathcal{K}_{\rm{N}}\two(\bm{k}_{1}, \bm{k}_{2}, \bm{k}_{3})
&=& b_{2}+ b_{1}F_{2}({\k}_{1},{\k}_{2})+ 
b_{\mathcal{K}^2}\mathcal{K}_{2}({\k}_{1},{\k}_{2})
+ f\,G_{2}({\k}_{1},{\k}_{2})
\mu_3^{2}+ {\cal Z}_2({\k}_{1},{\k}_{2})\,,
\end{eqnarray}
Here we decompose each ${\k}$ with respect to ${\n}$; ${k}_M^i= {k}_{M \|}n^i + {k}_{\bot M}^i = \mu_M k_M n^i + {k}_{\bot M}^i$, where ${k}_{M \|}$ is the parallel component ${k}_{M \|} = {k}_M^i {n}_i = \mu_{M} k_{M}$ and ${k}_{\bot M}^i$ is the transverse component ${k}_{\bot M}^i {n}^i  =0$ and $\mu_{M} ={\hat{\k}_M\cdot{\n}}$.   ${\cal Z}_2$ is a collection of the second  order biased-dependent redshift space distortion terms \cite{Bernardeau:2001qr}
 \begin{eqnarray}
{\cal Z}_2(\bm{k}_1,\bm{k}_2)&=&(f\mu_3 k_3)\left[\frac{\mu_1}{k_1} \left(b_1 + f\mu^2_2\right) + \frac{\mu_2}{k_2}\left(b_1 + f\mu^2_1\right) \right]\,,
\end{eqnarray}
where ${\k}_3 \equiv {\k}_1 + {\k}_2$. Furthermore, we made use of the  Euler equation to relate the peculiar velocity to the matter density contrast
\begin{eqnarray}
v({\k}) &=& \frac{\HH f}{k^2}\left[ \delta_m\one({\k}) +
\int \frac{\d^3 k_1}{(2\pi)^3}\frac{\d^3 k_2}{(2\pi)^3}\delta_m({\k}_1) \delta_m({\k}_2)  G_2({\k}_1,{\k}_2)
(2\pi)^3\delta^D\left({\k}_1 + {\k}_2 - {\k}\right)\right]\,,
\end{eqnarray}
where  $G_2$  are the kernel for the dark matter density field and  the peculiar velocity kernel at second order respectively:
 \begin{eqnarray}
G_2({\k}_1,{\k}_2)&=& \frac{3}{7} +\frac{1}{2} \frac{{\k}_1\cdot {\k}_2}{k_1k_2}\left(\frac{k_1}{k_2}+\frac{k_2}{k_1}\right) +\frac{4}{7} \left(\frac{{\k}_1\cdot {\k}_2}{k_1 k_2}\right)^2\,,
 \end{eqnarray}
 The galaxy bispectrum is given by \cite{Smith:2007sb,Bernardeau:2001qr}
\begin{eqnarray}\label{eq:bispectrumg}
B_{g}( {\k}_{1},  {\k}_{2},  {\k}_{3}) = \mathcal{K}_{\rm{N}}\one({\k}_{1}) \mathcal{K}_{{\rm{N}}}\one({\k}_{2}) \mathcal{K}_{\rm{N}}\two({\k}_{1},  {\k}_{2},{\k}_{3}) P_{m}(k_{1})P_{m}(k_{2}) +   \text{2 cy. p.} 
\end{eqnarray}
where $ P_{m}(k)$ is the matter power spectrum. $B_{g}$ depends on nine free parameters,  in the  real space limit, one can impose the homogeneity and  isotropy  of the triangular configurations to reduce the nine free parameters to three; magnitude of the two sides and the angle between them. In redshift, the map in equation \eqref{eq:distortedposition} introduces a unique line of sight which breaks isotropy. 
Using the closure relation (translation invariance) for the closed triangle  ${\k}_1+{\k}_2+{\k}_3 = 0$,  the number of free variables maybe reduced to six, this implies that we can fix $\mu_3$ using  $\mu_1 k_1 + \mu_2 k_2 +\mu_3 k_3 = 0$ and  $k_3$ in terms of $k_1$, $k_2$ and the angle between them $\mu_{12}$.   Using the trigonometric identity,  $\mu_{2}$ may be expressed in terms of  $\mu_{1}$ and the azimuthal angle $\phi_n$
\begin{eqnarray}\label{eq:angles}
\mu_{2} &=& \mu_{1}\mu_{12} \pm \sqrt{1 - \mu_{1}^{2} }\sqrt{1 - \mu_{12}^{2}}\cos \phi_n\,,
\\
\mu_3 &=& - \frac{k_1}{k_3} \mu_1 - \frac{k_2}{k_3} \mu_2 \,.
\end{eqnarray}
where  $\mu_{12} ={\hat{\k}_1\cdot\hat{\k}_2}$ is the angle between ${\k}_1$ and ${\k}_2$.  This  decomposition which reduces the number of free parameters from nine to five, is the most optimal decomposition of the galaxy bispectrum in redshift space, $B_g({k}_1,{k}_2,\mu_{12},\mu_{1},\phi_n)$ and was first introduced in \cite{Scoccimarro:1999ed,Scoccimarro:2000sn}.

We shall go slightly higher to reduce the number of free parameters from five to four by averaging over the dependence on azimuthal angle to obtain the so-called $\phi_{n}$-average galaxy bispectrum
\begin{eqnarray}
B_{g }^{\phi_n}({k}_1,{k}_2,\mu_{12},\mu_{1}) \equiv \frac{1}{2\pi}\int_{0}^{2\pi} \d \phi_nB_g({k}_1,{k}_2,\mu_{12},\mu_1,\phi_n) \,
\end{eqnarray}
It helps to reduce the dimensionality of the data structures of the bispectrum measurements.  For a fixed cosmological model, the loss of information has been shown to be very small \cite{Gagrani:2016rfy}. 
 In this limit, the spherical harmonics ${Y_{L }^{M}}$ reduces to the Legendre polynominial. Expanding $\mu_{12}$ in Legendre polynomial as well leads to 
\begin{eqnarray}\label{eq:Bgmultipoles}
B_{g}^{\phi_n}({k}_1,{k}_2,\mu_{12},\mu_{1}) &=& \sum_{L = 0}^{\infty}
\sum_{\ell_{12}}B^{\phi_n}_{gL \ell_{12}}({k}_1,{k}_2)
\mathcal{L}_{\ell_{12}}(\mu_{12}) \mathcal{L}_{L}(\mu_1)\,, 
\end{eqnarray}
  Using the orthogonality condition we obtain the multipole moments with respect to $L$ and $\ell_{12}$ of the galaxy bispectrum 
\begin{eqnarray}\label{eq:Bmultipoles1}
B_{gL \ell_{12}}^{\rm{N}\phi_n}({k}_1,{k}_2) &=& {{(2\ell_{12}+1)\over 2}} {{(2L+1)\over 2}}\int_{-1}^{1}\d\mu_{1}\int_{-1}^{1}\d\mu_{12} \,
\\ \nonumber && \qquad  \qquad  ~~~~~~~ \times ~
B_g^{\rm{N}\phi_n}({k}_1,{k}_2,\mu_{12},\mu_{1}) \mathcal{L}_{\ell_{12}}(\mu_{12}) \mathcal{L}_{L}(\mu_1)\,.
\end{eqnarray}
If one chooses to count the number of triangles instead of decomposing the angle between ${\k}_1$ and ${\k}_2$ into multipoles, summing over the first few multipoles of order  $\ell_{12}$ will recover $B_{g L \lambda}$ provided $B_g$ is a well behaved function of its other argurments.

\section{More details on the technical derivation}\label{sec:detials}

\subsection{Details on the derivation of 3PCF }

We use the spherical harmonics addition theorem 
\begin{eqnarray}\label{eq:orthogonality}
 \mathcal{L}_{\ell} (\nu_{13})=\frac{4\pi}{(2\ell +1)} \sum_{m=-\ell}^{\ell }Y^{\ast}_{\ell m}(\hat{\r}_{13})Y_{\ell m}(\hat{\r}_{23})=\frac{4\pi}{(2\ell +1)} \sum_{m=-\ell}^{\ell }Y_{\ell m}(\hat{\r}_{13})Y^{\ast}_{\ell m}(\hat{\r}_{23})
\end{eqnarray}
to express  the Legendre polynomials in the spherical harmonics basis. The product of spherical harmonics  is given by
\begin{eqnarray}
Y_{\ell_1,m_1}({\r}_{13}) 
Y_{\ell_2,m_2}({\r}_{13})
&=&\sum_{m_3=-\ell_3}^{m_3=\ell_3}\sum_{\ell_3=|\ell_1-\ell_2|}^{\ell_1+\ell_2}\left( {\begin{array}{ccc}
\ell_1& \ell_2&  \ell_3\\
  m_1& {m_2}&m_3  \\
 \end{array} } \right)\mathcal{H}_{\ell_1,\ell_2, \ell_3}^{0,0,0}Y_{\ell_3 m_3}({\r}_{13})\,,
\label{eq:addsph}
\end{eqnarray}
where we have separated $\mathcal{G}_{\ell_1\ell_2\ell_3}^{m_1m_2m} $ into  m-dependent 3j symbol and 
 \begin{eqnarray}
\mathcal{H}_{\ell_{1} \ell_{2} \ell_3}^{0,0,0}&=&\sqrt{\frac{(2\ell_1+1)(2\ell_{2}+1)(2\ell_3+1)}{4\pi}}
\begin{pmatrix}
 \ell_{1}  & \ell_{2} & \ell_{3}\\
  0 & 0 & 0
\end{pmatrix} 
\,.
\end{eqnarray}
Putting the multipole expansion of $B_{g}^{\phi_n}$  in equation \eqref{eq:3pcf2} gives
\begin{eqnarray}\nonumber 
\zeta_g({\r}_{13},{\r}_{23}, {\n}) &=&\sum_{\ell_1,\ell_2 L\ell_{12}} {(2\ell_1+1)(2\ell_2+1)} i^{\ell_1+\ell_2}\int \frac{\d k_1 k^2_1}{(2\pi)^3}
\int \frac{\d k_2 k^2_2}{(2\pi)^3}
\int\d {\k}_{{\bot}_{1}}\int \d {\k}_{{\bot}_{2}}j_{\ell_1}(k_1 r_{13})
\\  &&
 \times j_{\ell_2}(k_2 r_{23})  B^{\phi_n}_{gL \ell_{12}}({k}_1,{k}_2)
\mathcal{L}_{\ell_{12}}(\hat{\k}_1\cdot\hat{\k}_{2}) \mathcal{L}_{L}({\hat{\k}}_{1}\cdot {\n})\
\mathcal{L}_{\ell_1}({\hat{\r}}_{13}\cdot{\hat{\k}}_{1})\mathcal{L}_{\ell_2}({\hat{\r}}_{23}\cdot{\hat{\k}}_{2})\,,
\label{eq:3pcf3}
\end{eqnarray}
We  made use of equation \eqref{eq:orthogonality} to express the Legendre polynomial in terms of the spherical harmonics
\begin{eqnarray}
&&\mathcal{L}_{\ell_{12}} ({\k}_1\cdot{\k}_2) \mathcal{L}_{L}({\k}_1\cdot{\n})\mathcal{L}_{\ell_1}({{\r}_{13}}\cdot{{\k}_{1}})\mathcal{L}_{\ell_2}({{\r}_{23}}\cdot{{\k}_{2}})\\  \nonumber 
&=& \frac{4\pi}{(2\ell_1+1)}\frac{4\pi}{(2\ell_2+1)}\frac{4\pi}{(2\ell_{12}+1)}\frac{4\pi}{(2L+1)}
\sum_{m=-\ell_1}^{\ell_1}  \sum_{m=-\ell_2}^{\ell_2} \sum_{m=-\ell_{12}}^{\ell_{12}}\sum_{m=-L}^{L}
\\ \nonumber && \times
Y_{\ell_1 m_1}^{\ast}({{\r}}_{13})
Y_{\ell_2 m_2}({{\r}}_{23})Y_{L M}({\n})
Y_{\ell_1 m_1}({\k}_1)Y^{\ast}_{\ell_{12} m_{12}}({\k}_1)Y^{\ast}_{L M}({\k}_{1})
Y^{\ast}_{\ell_2 m_2}({\k}_2)Y_{\ell_{12} m_{12}}({\k}_2)\,.
\end{eqnarray}
Then  performed the $\hat{\k}_{2}$  and  $\hat{\k}_{1}$  using
\begin{eqnarray}\label{orthogonality}
\int d \Omega_{k_2}  Y_{\ell_2 m_2} ({\k}_2)Y^*_{\ell_{12} m_{12}} ({\k}_2) &=& \delta_{\ell_2 \ell_{12}} \delta_{m_2 m_{12}} \,,
\\ 
\quad \int d^2{{ \hat{\k}_{1}}}  Y^{\star}_{\ell_1 m_1}({{\hat  {\k}_{1}}})
 Y^{\star}_{\ell_{12}m_{12}}({{\hat {\k}_{1}}})Y_{L M}({{\hat
 {\k}_{1}}}) 
& =&\begin{pmatrix}
 \ell_{12}  & L & \ell_1\\
 m_{12}  & M&  m_1
\end{pmatrix} \mathcal{H}_{\ell_{12} L \ell_1}^{0,0,0} \,.
\label{eq:gauntintegral}
\end{eqnarray}
After performing the Kronecker delta summation and some algebraic simplification we find
\begin{eqnarray}\nonumber
\zeta_g({\r}_{13},{\r}_{23} ,{\n}) &=&\sum_{{L} \ell_1 \ell_{12}}\sum_{m_1 m_{12} M} \frac{4\pi}{(2\ell_{12}+1)}\frac{4\pi}{(2L+1)} \int \frac{\d k_1 k^2_1}{2\pi^2}
\int \frac{\d k_2 k^2_2}{2\pi^2}
i^{\ell_1+\ell_{12}}\mathcal{G}_{\ell_1\ell_{12}L}^{m_1m_{12} M} 
\\ &&
\times B_{\text{g}\ell_{12} L }^{\phi_{n}}(z,k_1,k_2)
j_{\ell_1}(k_1 r_{13})j_{\ell_{12}}(k_2 r_{23}) Y_{\ell_1 m_1}^{\ast}({\hat{\r}}_{13})
Y_{\ell_{12} m_{12}}(\hat{\bf{r}}_{23})Y_{L M}({\n})\,,
\label{eq:3pcf4}
\end{eqnarray}
Substituting equation \eqref{eq:3pcf4} in equation \eqref{eq:zetamultipoles} gives
\begin{eqnarray}
{\zeta}^{\phi_n}_{L\ell_{12}}({r}_{13},{r}_{23}) &=&
i^{\ell_3+\ell_{12}}
\int_{0}^{\infty} \frac{\d k_1 k^2_1}{2\pi^2}
\int_{0}^{\infty}\frac{\d k_2 k^2_2}{2\pi^2}B_{{g}L\ell_{12}  }(k_1,k_2)
 j_{\ell_3}(k_1 r_{12})j_{\ell_{12}}(k_2 r_{23})\,.
\end{eqnarray}

\subsection{Details on the derivation of 3PCF covariance}

We make use of the plane wave approximation
   \begin{eqnarray}
e^{i{\k}_1\cdot {\r}_{13}} & =&\sum_{\ell_3 }(2\ell_3+1) i^{\ell_3} j_{\ell_3} (kr_{13})
 \mathcal{L}_{\ell_3}(\hat{\bf{r}}_{13}\cdot \hat{{\k}_1}) \,,
 \\
e^{i{\k}_2\cdot {\r}_{23}} & =&\sum_{\ell_3 }(2\ell_3+1) i^{\ell_3} 
 j_{\ell_3}(k_2r_{23})
 \mathcal{L}_{\ell_3}(\hat{{\r}}_{23}\cdot \hat{\k}_2)\,.
\end{eqnarray}
Substituting these in equation \eqref{eq:beforeplaneapprox} and performing some algebraic simplification leads to 
  \begin{eqnarray}\nonumber 
  {\rm{Cov}}\left[ \hat{\zeta}_g({\r}_{13},{\r}_{23},{\n})\hat{\zeta}_g({\r}'_{13},{\r}'_{23},{\n})\right] 
  & =&   \int {\d k_1} k_1^2 \int {\d k_2}  k_2^2  \int \d^2 {\k}_1 \int \d^2 {\k}_2   \sum_{L_2 \ell_{12}}      \frac{\left(P({k}_1)P({k}_2)P({k}_{12})\right)_{L _2\ell_{12}}  }{(2\pi)^6V_{s}}
 \\ \nonumber &&
   \sum_{\ell_1\ell_1' }\sum_{\ell_2\ell_2' } (2\ell_1+1)(2\ell_1'+1) 
    (2\ell_2+1)(2\ell_2'+1)(i)^{\ell_1 + \ell_2}(-i)^{\ell_1' +\ell_2'} 
   \\ \nonumber &&
   \bigg[ j_ {\ell_1}(k_1r_{13})
 j_{\ell_1'}(k_1r'_{13}) j_ {\ell_2}(k_2r_{23})
 j_{\ell_2'}(k_2r'_{23})
  \\ \nonumber && \times
  X^1_{\ell_1\ell_1'\ell_2\ell_2'L_2\ell_{12}}([\hat{\k}_1,\hat{\r}_{13},\hat{\r}'_{13}],[\hat{\k}_2,\hat{\r}_{23},\hat{\r}'_{23}],{\n}) 
 \\ \nonumber &&
 +  j_ {\ell_1}(k_1r_{13})
 j_{\ell_1'}(k_1r'_{23}) j_ {\ell_2}(k_2r_{23})
 j_{\ell_2'}(k_2r'_{13})
   \\ && \times
X^2_{\ell_1\ell_1'\ell_2\ell_2'L_2\ell_{12}}([\hat{\k}_1,\hat{\r}_{13},\hat{\r}'_{23}],[\hat{\k}_2,\hat{\r}_{23},\hat{\r}'_{13}],{\n}) 
 \bigg] \,,
 \label{eq:3pcfocovarianceplanewave}
\end{eqnarray}
where 
\begin{eqnarray}
X^1_{\ell_1\ell_1'\ell_2\ell_2'L\ell_{12}}([\hat{\k}_1,\hat{\r}_{13},\hat{\r}'_{13}],[\hat{\k}_2,\hat{\r}_{23},\hat{\r}'_{23}],{\n}) &=&\mathcal{L}_{\ell_1}(\hat{{\r}}_{13}\cdot \hat{{\k}_1})\mathcal{L}_{\ell_1'}(\hat{{\r}'}_{13}\cdot \hat{{\k}}_1)\mathcal{L}_{\ell_2}(\hat{{\r}}_{23}\cdot \hat{\k}_2)
 \\ \nonumber &&
\mathcal{L}_{\ell_2'}(\hat{\r}'_{23}\cdot \hat{\k}_2)\mathcal{L}_{L}({\k}_1\cdot{\n}) \mathcal{L}_{\ell_{12}}({\k}_{1}\cdot{\k}_{2})\,,
\\ 
X^2_{\ell_1\ell_1'\ell_2\ell_2'L_2\ell_{12}}([\hat{\k}_1,\hat{\r}_{13},\hat{\r}'_{23}],[\hat{\k}_2,\hat{\r}_{23},\hat{\r}'_{13}],{\n})   &=& \mathcal{L}_{\ell_1}(\hat{{\r}}_{13}\cdot \hat{{\k}_1})\mathcal{L}_{\ell_1'}(\hat{{\r}'}_{23}\cdot \hat{{\k}_1})
 \mathcal{L}_{\ell_2}(\hat{{\r}}_{23}\cdot \hat{{\k}_2})
  \\ \nonumber &&
 \mathcal{L}_{\ell_2'}(\hat{{\r}'}_{13}\cdot \hat{{\k}_2})\mathcal{L}_{L_2}({\k}_1\cdot{\n}) \mathcal{L}_{\ell_{12}}({\k}_{1}\cdot{\k}_{2})
 \,.
\end{eqnarray}

Putting equation \eqref{eq:3pcfocovarianceplanewave} in equation \eqref{eq:multipolecovarincedef} leads to 

 \begin{eqnarray}\nonumber
 {\rm{Cov}}\left[\hat{\zeta}^{\phi_n}_{\ell_{13}L}({r}_{13},{r}_{23}) \hat{\zeta}^{\phi_n}_{\ell'_{13}L'}({r}'_{13},{r}'_{23})\right] 
&=&
\bigg[ \frac{(2L+1)(2\ell_{13} +1)}{\mathcal{L}_{\ell_{12},L, \ell_3}} \frac{{{(2L'+1)}}(2\ell'_{13} +1)}{\mathcal{L}_{\ell'_{13},L', \ell'_3}}   \bigg]
 \\ \nonumber  &&
 \sum_{m_{13}Mm_3}
\left( {\begin{array}{ccc}
\ell_{13}& L&  \ell_3\\
  m_{13}& {M}&m_3  \\
 \end{array} } \right)
 \sum_{m'_{13}M'm'_3}
\left( {\begin{array}{ccc}
\ell'_{13}& L'&  \ell'_3\\
  m'_{13}& {M'}&m'_3  \\
 \end{array} } \right)
 \\ \nonumber &&
\int {\d k_1} k_1^2 \int {\d k_2}  k_2^2     \sum_{L_2 \ell_{12}}      \frac{\left(P({k}_1)P({k}_2)P({k}_{12})\right)_{L _2\ell_{12}}}{  (V_{s})2\pi)^6} 
 \\ \nonumber &&
   \sum_{\ell_1\ell_1' }\sum_{\ell_2\ell_2' } (2\ell_1+1)(2\ell_1'+1) 
    (2\ell_2+1)(2\ell_2'+1)(i)^{\ell_1 + \ell_2}(-i)^{\ell_1' +\ell_2'}
       \\ \nonumber &&
  \int\mathrm{d}^2\mathbf{r}_{13}
\int\mathrm{d}^2\mathbf{r}_{23}\int\mathrm{d}^2\mathbf{r}'_{13}
\int\mathrm{d}^2\mathbf{r}'_{23}
\int \d^2 {\n}     \int \d^2 {\k}_1 \int \d^2 {\k}_2
   \\ \nonumber &&
   \bigg[ j_ {\ell_1}(k_1r_{13})
 j_{\ell_1'}(k_1r'_{13}) j_ {\ell_2}(k_2r_{23})
 j_{\ell_2'}(k_2r'_{23}) 
  \\ \nonumber &&
 X^1_{\ell_1\ell_1'\ell_2\ell_2'L_2\ell_{12}}([\hat{\k}_1,\hat{\r}_{13},\hat{\r}'_{13}],[\hat{\k}_2,\hat{\r}_{23},\hat{\r}'_{23}],{\n}) 
 \\ \nonumber &&
 +  j_ {\ell_1}(k_1r_{13})
 j_{\ell_1'}(k_1r'_{23}) j_ {\ell_2}(k_2r_{23})
 j_{\ell_2'}(k_2r'_{13})
  \\ \nonumber &&
X^2_{\ell_1\ell_1'\ell_2\ell_2'L_2\ell_{12}}([\hat{\k}_1,\hat{\r}_{13},\hat{\r}'_{23}],[\hat{\k}_2,\hat{\r}_{23},\hat{\r}'_{13}],{\n}) 
   \bigg]
     \\ \nonumber && \times
       \mathcal{Y}_{\ell_{3} \ell_{13} L}^{m_{3} m_{13} M}(\hat{\r}_{13},\hat{\r}_{23},{\n}) 
   \mathcal{Y}_{\ell'_{3} \ell'_{13} L'}^{m'_{3} m'_{13} M'}(\hat{\r}'_{13},\hat{\r}'_{23},{\n}') \,,
 \end{eqnarray}
 where
 \begin{eqnarray}
 \mathcal{Y}_{\ell_{3} \ell_{13} L}^{m_{3} m_{13} M}(\hat{\r}_{13},\hat{\r}_{23},{\n}) &=& \left[ Y^{\ast}_{\ell_3 m_3}(\hat{\r}_{13})Y^{\ast}_{\ell_{13} m_{13}}(\hat{\r}_{23})
Y_{LM}({\n}) \right]\,.
 \end{eqnarray}
 
 Using the addition theorem, we can express the Legendre polynomial in terms of the spherical harmonics
 \begin{eqnarray} \nonumber
 X^1_{\ell_1\ell_1'\ell_2\ell_2'L\ell_{12}}([\hat{\k}_1,\hat{\r}_{13},\hat{\r}'_{13}],[\hat{\k}_2,\hat{\r}_{23},\hat{\r}'_{23}],{\n}) &=&
\frac{4\pi}{(2\ell_1+1)}\frac{4\pi}{(2\ell_1'+1)}\frac{4\pi}{(2\ell_2+1)}\frac{4\pi}{(2\ell_2'+1)}
\frac{4\pi}{(2L_2+1)}\frac{4\pi}{(2\ell_{12}+1)}
 \\ \nonumber &&
\sum_{m_1=-\ell_1}^{\ell_1}\sum_{m_1'=-\ell_1}^{\ell_1'} \sum_{m_2=-\ell_2}^{\ell_2}\sum_{m_2'=-\ell_2}^{\ell_2'} \sum_{m_{12}=-\ell_{12}}^{\ell_{12}}\sum_{M_2=-L_2}^{L_2} 
 \\ \nonumber &&
Y_{\ell_1 m_1}(\hat{\r}_{13})
Y_{\ell_1' m_1'}(\hat{\r}'_{13})
\\ \nonumber &&
Y_{\ell_2 m_2}\hat({\r}_{23})
Y_{\ell_2' m_2'}\hat({\r}'_{23})
\\ \nonumber &&
Y^{\ast}_{\ell_{12} m_{12}}(\hat{\k}_{1})Y_{L_2M_2}(\hat{\k}_1)Y^{\ast}_{\ell_1 m_1}(\hat{\k}_{1})Y^{\ast}_{\ell_1' m_1'}(\hat{\k}_{1})
\\ &&
Y^{\ast}_{\ell_2' m_2'}(\hat{\k}_{2})Y^{\ast}_{\ell_2 m_2}(\hat{\k}_{2})Y_{\ell_{12} m_{12}}(\hat{\k}_2)
\\ \nonumber &&
Y^{\ast}_{L_2 M_2 }({\n}) \,.
\label{eq:additiontheoremexpanded}
 \end{eqnarray}
Note that $Y_{\ell m}^{\ast} = (-1)^{m} Y_{\ell -m}$. Using equation \eqref{eq:additiontheoremexpanded}, we can not perform the angular integrals using   equation \eqref{orthogonality}, \eqref{eq:gauntintegral} and 
\begin{eqnarray}
&& \int \d^2 {\k}_1Y^{\ast}_{\ell_{12} m_{12}}({\k}_{1})Y_{L_2M_2}({\k}_1)Y^{\ast}_{\ell_1 m_1}({\k}_{1})Y^{\ast}_{\ell_1' m_1'}({\k}_{1})
\\ \nonumber
&=&\sum_{M_3=-L_3}^{M_3=L_3}\sum_{L_3=|\ell_1-\ell_1'|}^{\ell_1+\ell_1'}
\begin{pmatrix}
 L_2 & \ell_{12} &L_3\\
M_2 & m_{12} & M_3
\end{pmatrix}
\begin{pmatrix}
\ell_1& \ell'_{1} &L_3\\
m_1 & m'_{1} & M_3
\end{pmatrix}
\mathcal{L}_{L_2 \ell_{12} L_3}^{0,0,0}
\mathcal{L}_{\ell_1 \ell'_{1} L_3}^{0,0,0}
\,.
\end{eqnarray}
Performing $\ell_1$ and $\ell_2$ sums  and $\ell'_1$ and $\ell'_2$ sums and  using the following definition of the 9j symbol
\begin{eqnarray}
&& \sum_{m_{13}Mm_3} \sum_{m'_{13}M'm'_3}
\sum_{m_{12}M_2M_3 }
\left( {\begin{array}{ccc}
\ell_{13}& L&  \ell_3\\
  m_{13}& M&m_3  \\
 \end{array} } \right)
\left( {\begin{array}{ccc}
\ell'_{13}& L'&  \ell'_3\\
  m'_{13}& M'&m'_3  \\
 \end{array} } \right)\begin{pmatrix}
 \ell_{12}&L_2 &L_3\\
 m_{12} &M_2 & M_3
\end{pmatrix}
\\ \nonumber &&
 \begin{pmatrix}
  \ell_3 & \ell'_{3} &L_3\\
  m_3 & m'_{3} & M_3
\end{pmatrix}
\begin{pmatrix}
  \ell_{13} & \ell'_{13} & \ell_{12}\\
  m_{13} & m'_{13} & m_{12}
\end{pmatrix}
\begin{pmatrix}
L &L' & L_{2}\\
  M& M' & M_{2}
\end{pmatrix}
 = \begin{Bmatrix}
   \ell_{13}& L&  \ell_3\\
 \ell'_{13}& L'&  \ell'_3\\
 \ell_{12}&L_2 &L_3
  \end{Bmatrix}
  \,,
\end{eqnarray}
lead to equation \eqref{eq:covariance3pctbody1}. Note that simplification of 
\begin{eqnarray}
X^2_{\ell_1\ell_1'\ell_2\ell_2'L_2\ell_{12}}([{\k}_1,{\r}_{13},{\r}'_{23}],[{\k}_2,{\r}_{23},{\r}'_{13}],{\n})   &=& \mathcal{L}_{\ell_1}(\hat{{\r}}_{13}\cdot \hat{{\k}_1})\mathcal{L}_{\ell_1'}(\hat{{\r}'}_{23}\cdot \hat{{\k}_1})
 \mathcal{L}_{\ell_2}(\hat{{\r}}_{23}\cdot \hat{{\k}_2})
  \\ \nonumber &&
 \mathcal{L}_{\ell_2'}(\hat{{\r}'}_{13}\cdot \hat{{\k}_2})\mathcal{L}_{L_2}({\k}_1\cdot{\n}) \mathcal{L}_{\ell_{12}}({\k}_{1}\cdot{\k}_{2})\,,
\end{eqnarray}
follows the same procedure.




\begin{thebibliography}{10}

\bibitem{Gil-Marin:2014sta}
H.~Gil-Marín, J.~Noreña, L.~Verde, W.~J. Percival, C.~Wagner, M.~Manera, and
  D.~P. Schneider, {\it {The power spectrum and bispectrum of SDSS DR11 BOSS
  galaxies -- I. Bias and gravity}},  {\em Mon. Not. Roy. Astron. Soc.} {\bf
  451} (2015), no.~1 539--580, [\href{http://arxiv.org/abs/1407.5668}{{\tt
  arXiv:1407.5668}}].

\bibitem{Bartolo:2004if}
N.~Bartolo, E.~Komatsu, S.~Matarrese, and A.~Riotto, {\it {Non-Gaussianity from
  inflation: Theory and observations}},  {\em Phys. Rept.} {\bf 402} (2004)
  103--266, [\href{http://arxiv.org/abs/astro-ph/0406398}{{\tt
  astro-ph/0406398}}].

\bibitem{Creminelli:2004pv}
P.~Creminelli and M.~Zaldarriaga, {\it {CMB 3-point functions generated by
  non-linearities at recombination}},  {\em Phys. Rev.} {\bf D70} (2004)
  083532, [\href{http://arxiv.org/abs/astro-ph/0405428}{{\tt
  astro-ph/0405428}}].

\bibitem{Creminelli:2013nua}
P.~Creminelli, J.~Gleyzes, L.~Hui, M.~Simonovi, and F.~Vernizzi, {\it
  {Single-Field Consistency Relations of Large Scale Structure. Part III: Test
  of the Equivalence Principle}},  {\em JCAP} {\bf 1406} (2014) 009,
  [\href{http://arxiv.org/abs/1312.6074}{{\tt arXiv:1312.6074}}].

\bibitem{Chen:2006nt}
X.~Chen, M.-x. Huang, S.~Kachru, and G.~Shiu, {\it {Observational signatures
  and non-Gaussianities of general single field inflation}},  {\em JCAP} {\bf
  01} (2007) 002, [\href{http://arxiv.org/abs/hep-th/0605045}{{\tt
  hep-th/0605045}}].

\bibitem{Scoccimarro:2011pz}
R.~Scoccimarro, L.~Hui, M.~Manera, and K.~C. Chan, {\it {Large-scale Bias and
  Efficient Generation of Initial Conditions for Non-Local Primordial
  Non-Gaussianity}},  {\em Phys.Rev.} {\bf D85} (2012) 083002,
  [\href{http://arxiv.org/abs/1108.5512}{{\tt arXiv:1108.5512}}].

\bibitem{Arkani-Hamed:2015bza}
N.~Arkani-Hamed and J.~Maldacena, {\it {Cosmological Collider Physics}},
  \href{http://arxiv.org/abs/1503.08043}{{\tt arXiv:1503.08043}}.

\bibitem{Slepian:2015qza}
Z.~Slepian and D.~J. Eisenstein, {\it {Computing the three-point correlation
  function of galaxies in $\mathcal {O}(N^2)$ time}},  {\em Mon. Not. Roy.
  Astron. Soc.} {\bf 454} (2015), no.~4 4142--4158,
  [\href{http://arxiv.org/abs/1506.02040}{{\tt arXiv:1506.02040}}].

\bibitem{Szapudi:2004gg}
I.~Szapudi, {\it {Three - point statistics from a new perspective}},  {\em
  Astrophys. J.} {\bf 605} (2004) L89,
  [\href{http://arxiv.org/abs/astro-ph/0404476}{{\tt astro-ph/0404476}}].

\bibitem{Slepian:2016kfz}
Z.~Slepian et~al., {\it {Detection of baryon acoustic oscillation features in
  the large-scale three-point correlation function of SDSS BOSS DR12 CMASS
  galaxies}},  {\em Mon. Not. Roy. Astron. Soc.} {\bf 469} (2017), no.~2
  1738--1751, [\href{http://arxiv.org/abs/1607.06097}{{\tt arXiv:1607.06097}}].

\bibitem{Slepian:2016weg}
Z.~Slepian and D.~J. Eisenstein, {\it {Modelling the large-scale redshift-space
  3-point correlation function of galaxies}},  {\em Mon. Not. Roy. Astron.
  Soc.} {\bf 469} (2017), no.~2 2059--2076,
  [\href{http://arxiv.org/abs/1607.03109}{{\tt arXiv:1607.03109}}].

\bibitem{Slepian:2017lpm}
Z.~Slepian and D.~J. Eisenstein, {\it {A practical computational method for the
  anisotropic redshift-space three-point correlation function}},  {\em Mon.
  Not. Roy. Astron. Soc.} {\bf 478} (2018), no.~2 1468--1483,
  [\href{http://arxiv.org/abs/1709.10150}{{\tt arXiv:1709.10150}}].

\bibitem{Friesen:2017acf}
B.~Friesen et~al., {\it {Galactos: Computing the Anisotropic 3-Point
  Correlation Function for 2 Billion Galaxies}},
  \href{http://arxiv.org/abs/1709.00086}{{\tt arXiv:1709.00086}}.

\bibitem{Scoccimarro:1997st}
R.~Scoccimarro, S.~Colombi, J.~N. Fry, J.~A. Frieman, E.~Hivon, et~al., {\it
  {Nonlinear evolution of the bispectrum of cosmological perturbations}},  {\em
  Astrophys.J.} {\bf 496} (1998) 586,
  [\href{http://arxiv.org/abs/astro-ph/9704075}{{\tt astro-ph/9704075}}].

\bibitem{Scoccimarro:2000sn}
R.~Scoccimarro, {\it {The bispectrum: from theory to observations}},  {\em
  Astrophys. J.} {\bf 544} (2000) 597,
  [\href{http://arxiv.org/abs/astro-ph/0004086}{{\tt astro-ph/0004086}}].

\bibitem{Smith:2007sb}
R.~E. Smith, R.~K. Sheth, and R.~Scoccimarro, {\it {An analytic model for the
  bispectrum of galaxies in redshift space}},  {\em Phys. Rev.} {\bf D78}
  (2008) 023523, [\href{http://arxiv.org/abs/0712.0017}{{\tt
  arXiv:0712.0017}}].

\bibitem{Sugiyama:2018yzo}
N.~S. Sugiyama, S.~Saito, F.~Beutler, and H.-J. Seo, {\it {A complete FFT-based
  decomposition formalism for the redshift-space bispectrum}},  {\em Mon. Not.
  Roy. Astron. Soc.} {\bf 484} (2019), no.~1 364--384,
  [\href{http://arxiv.org/abs/1803.02132}{{\tt arXiv:1803.02132}}].

\bibitem{Varshalovich:book}
D.~A. Varshalovich, A.~N. Moskalev, and V.~K. Khersonskii, {\em Quantum Theory
  of Angular Momentum}.
\newblock WORLD SCIENTIFIC, 1988.

\bibitem{Fang:2020vhc}
X.~Fang, T.~Eifler, and E.~Krause, {\it {2D-FFTLog: Efficient computation of
  real space covariance matrices for galaxy clustering and weak lensing}},
  \href{http://arxiv.org/abs/2004.04833}{{\tt arXiv:2004.04833}}.

\bibitem{Hamilton:1999uv}
A.~Hamilton, {\it {Uncorrelated modes of the nonlinear power spectrum}},  {\em
  Mon. Not. Roy. Astron. Soc.} {\bf 312} (2000) 257--284,
  [\href{http://arxiv.org/abs/astro-ph/9905191}{{\tt astro-ph/9905191}}].

\bibitem{Hamilton:2015ascl.soft12017H}
A.~J.~S. {Hamilton}, {\it {FFTLog: Fast Fourier or Hankel transform}},  Dec.,
  2015.

\bibitem{Wang:2016wjr}
{\bf BOSS} Collaboration, Y.~Wang et~al., {\it {The clustering of galaxies in
  the completed SDSS-III Baryon Oscillation Spectroscopic Survey: tomographic
  BAO analysis of DR12 combined sample in configuration space}},  {\em Mon.
  Not. Roy. Astron. Soc.} {\bf 469} (2017), no.~3 3762--3774,
  [\href{http://arxiv.org/abs/1607.03154}{{\tt arXiv:1607.03154}}].

\bibitem{Zarrouk:2018vwy}
P.~Zarrouk et~al., {\it {The clustering of the SDSS-IV extended Baryon
  Oscillation Spectroscopic Survey DR14 quasar sample: measurement of the
  growth rate of structure from the anisotropic correlation function between
  redshift 0.8 and 2.2}},  {\em Mon. Not. Roy. Astron. Soc.} {\bf 477} (2018),
  no.~2 1639--1663, [\href{http://arxiv.org/abs/1801.03062}{{\tt
  arXiv:1801.03062}}].

\bibitem{Ross:2016gvb}
{\bf BOSS} Collaboration, A.~J. Ross et~al., {\it {The clustering of galaxies
  in the completed SDSS-III Baryon Oscillation Spectroscopic Survey:
  Observational systematics and baryon acoustic oscillations in the correlation
  function}},  {\em Mon. Not. Roy. Astron. Soc.} {\bf 464} (2017), no.~1
  1168--1191, [\href{http://arxiv.org/abs/1607.03145}{{\tt arXiv:1607.03145}}].

\bibitem{Blanchard:2019oqi}
{\bf Euclid} Collaboration, A.~Blanchard et~al., {\it {Euclid preparation: VII.
  Forecast validation for Euclid cosmological probes}},  {\em Astron.
  Astrophys.} {\bf 642} (2020) A191,
  [\href{http://arxiv.org/abs/1910.09273}{{\tt arXiv:1910.09273}}].

\bibitem{Aghamousa:2016zmz}
{\bf DESI} Collaboration, A.~Aghamousa et~al., {\it {The DESI Experiment Part
  I: Science,Targeting, and Survey Design}},
  \href{http://arxiv.org/abs/1611.00036}{{\tt arXiv:1611.00036}}.

\bibitem{Santos:2015gra}
M.~G. Santos et~al., {\it {Cosmology with a SKA HI intensity mapping survey}},
  \href{http://arxiv.org/abs/1501.03989}{{\tt arXiv:1501.03989}}.

\bibitem{Ade:2015xua}
{\bf Planck} Collaboration, P.~A.~R. Ade et~al., {\it {Planck 2015 results.
  XIII. Cosmological parameters}},  {\em Astron. Astrophys.} {\bf 594} (2016)
  A13, [\href{http://arxiv.org/abs/1502.01589}{{\tt arXiv:1502.01589}}].

\bibitem{Aghanim:2018eyx}
{\bf Planck} Collaboration, N.~Aghanim et~al., {\it {Planck 2018 results. VI.
  Cosmological parameters}},  \href{http://arxiv.org/abs/1807.06209}{{\tt
  arXiv:1807.06209}}.

\bibitem{Ellis2009}
G.~F.~R. Ellis, {\it Republication of: Relativistic cosmology},  {\em General
  Relativity and Gravitation} {\bf 41} (Mar, 2009) 581--660.

\bibitem{Challinor:2011bk}
A.~Challinor and A.~Lewis, {\it {The linear power spectrum of observed source
  number counts}},  {\em Phys.Rev.} {\bf D84} (2011) 043516,
  [\href{http://arxiv.org/abs/1105.5292}{{\tt arXiv:1105.5292}}].

\bibitem{Alonso:2015uua}
D.~Alonso, P.~Bull, P.~G. Ferreira, R.~Maartens, and M.~Santos, {\it {Ultra
  large-scale cosmology in next-generation experiments with single tracers}},
  {\em Astrophys. J.} {\bf 814} (2015), no.~2 145,
  [\href{http://arxiv.org/abs/1505.07596}{{\tt arXiv:1505.07596}}].

\bibitem{Desjacques:2016bnm}
V.~Desjacques, D.~Jeong, and F.~Schmidt, {\it {Large-Scale Galaxy Bias}},  {\em
  Phys. Rept.} {\bf 733} (2018) 1--193,
  [\href{http://arxiv.org/abs/1611.09787}{{\tt arXiv:1611.09787}}].

\bibitem{Umeh:2019qyd}
O.~Umeh, K.~Koyama, R.~Maartens, F.~Schmidt, and C.~Clarkson, {\it {General
  relativistic effects in the galaxy bias at second order}},  {\em JCAP} {\bf
  1905} (2019), no.~05 020, [\href{http://arxiv.org/abs/1901.07460}{{\tt
  arXiv:1901.07460}}].

\bibitem{Yankelevich:2018uaz}
V.~Yankelevich and C.~Porciani, {\it {Cosmological information in the
  redshift-space bispectrum}},  {\em Mon. Not. Roy. Astron. Soc.} {\bf 483}
  (2019), no.~2 2078--2099, [\href{http://arxiv.org/abs/1807.07076}{{\tt
  arXiv:1807.07076}}].

\bibitem{Zheng:2004eh}
Z.~Zheng, {\it {Projected three - point correlation functions and galaxy
  bias}},  {\em Astrophys. J.} {\bf 614} (2004) 527--532,
  [\href{http://arxiv.org/abs/astro-ph/0405527}{{\tt astro-ph/0405527}}].

\bibitem{Bertacca:2014hwa}
D.~Bertacca, {\it {Observed galaxy number counts on the light cone up to second
  order: III. Magnification bias}},  {\em Class. Quant. Grav.} {\bf 32} (2015),
  no.~19 195011, [\href{http://arxiv.org/abs/1409.2024}{{\tt
  arXiv:1409.2024}}].

\bibitem{Yoo:2014sfa}
J.~Yoo and M.~Zaldarriaga, {\it {Beyond the Linear-Order Relativistic Effect in
  Galaxy Clustering: Second-Order Gauge-Invariant Formalism}},  {\em Phys.
  Rev.} {\bf D90} (2014), no.~2 023513,
  [\href{http://arxiv.org/abs/1406.4140}{{\tt arXiv:1406.4140}}].

\bibitem{DiDio:2014lka}
E.~Di~Dio, R.~Durrer, G.~Marozzi, and F.~Montanari, {\it {Galaxy number counts
  to second order and their bispectrum}},  {\em JCAP} {\bf 1412} (2014) 017,
  [\href{http://arxiv.org/abs/1407.0376}{{\tt arXiv:1407.0376}}]. [Erratum:
  JCAP1506,no.06,E01(2015)].

\bibitem{Umeh:2015gza}
O.~Umeh, R.~Maartens, and M.~Santos, {\it {Nonlinear modulation of the HI power
  spectrum on ultra-large scales. I}},  {\em JCAP} {\bf 1603} (2016), no.~03
  061, [\href{http://arxiv.org/abs/1509.03786}{{\tt arXiv:1509.03786}}].

\bibitem{Umeh:2016thy}
O.~Umeh, {\it {Imprint of non-linear effects on HI intensity mapping on large
  scales}},  {\em JCAP} {\bf 1706} (2017), no.~06 005,
  [\href{http://arxiv.org/abs/1611.04963}{{\tt arXiv:1611.04963}}].

\bibitem{Matsubara:1994ApJ}
T.~{Matsubara}, {\it {Peculiar Velocity Effect on Galaxy Correlation Functions
  in Nonlinear Clustering Regime}},  {\em \apj} {\bf 424} (Mar, 1994) 30.

\bibitem{Scoccimarro:1999ed}
R.~Scoccimarro, H.~Couchman, and J.~A. Frieman, {\it {The Bispectrum as a
  signature of gravitational instability in redshift-space}},  {\em
  Astrophys.J.} {\bf 517} (1999) 531--540,
  [\href{http://arxiv.org/abs/astro-ph/9808305}{{\tt astro-ph/9808305}}].

\bibitem{Bianchi:2015oia}
D.~Bianchi, H.~Gil-Marín, R.~Ruggeri, and W.~J. Percival, {\it {Measuring
  line-of-sight dependent Fourier-space clustering using FFTs}},  {\em Mon.
  Not. Roy. Astron. Soc.} {\bf 453} (2015), no.~1 L11--L15,
  [\href{http://arxiv.org/abs/1505.05341}{{\tt arXiv:1505.05341}}].

\bibitem{Gagrani:2016rfy}
P.~Gagrani and L.~Samushia, {\it {Information Content of the Angular Multipoles
  of Redshift-Space Galaxy Bispectrum}},  {\em Mon. Not. Roy. Astron. Soc.}
  {\bf 467} (2017), no.~1 928--935,
  [\href{http://arxiv.org/abs/1610.03488}{{\tt arXiv:1610.03488}}].

\bibitem{Garcia:2020per}
K.~Garcia and Z.~Slepian, {\it {Improving the Line of Sight for the Anisotropic
  3-Point Correlation Function of Galaxies: Centroid and Unit-Vector-Average
  Methods Scaling as $\mathcal{O}(N^2)$}},
  \href{http://arxiv.org/abs/2011.03503}{{\tt arXiv:2011.03503}}.

\bibitem{Sugiyama:2019ike}
N.~S. Sugiyama, S.~Saito, F.~Beutler, and H.-J. Seo, {\it {Perturbation theory
  approach to predict the covariance matrices of the galaxy power spectrum and
  bispectrum in redshift space}},  \href{http://arxiv.org/abs/1908.06234}{{\tt
  arXiv:1908.06234}}.

\bibitem{Lee:2020ebj}
H.~Lee and C.~Dvorkin, {\it {Cosmological Angular Trispectra and Non-Gaussian
  Covariance}},  {\em JCAP} {\bf 05} (2020) 044,
  [\href{http://arxiv.org/abs/2001.00584}{{\tt arXiv:2001.00584}}].

\bibitem{Assassi:2017lea}
V.~Assassi, M.~Simonovi\'c, and M.~Zaldarriaga, {\it {Efficient evaluation of
  angular power spectra and bispectra}},  {\em JCAP} {\bf 11} (2017) 054,
  [\href{http://arxiv.org/abs/1705.05022}{{\tt arXiv:1705.05022}}].

\bibitem{Alcock:1979mp}
C.~Alcock and B.~Paczynski, {\it {An evolution free test for non-zero
  cosmological constant}},  {\em Nature} {\bf 281} (1979) 358--359.

\bibitem{TALMAN197835}
J.~D. Talman, {\it Numerical fourier and bessel transforms in logarithmic
  variables},  {\em Journal of Computational Physics} {\bf 29} (1978), no.~1 35
  -- 48.

\bibitem{2009CoPhC.180..332T}
J.~D. {Talman}, {\it {NumSBT: A subroutine for calculating spherical Bessel
  transforms numerically}},  {\em Computer Physics Communications} {\bf 180}
  (Feb., 2009) 332--338.

\bibitem{McEwen:2016fjn}
J.~E. McEwen, X.~Fang, C.~M. Hirata, and J.~A. Blazek, {\it {FAST-PT: a novel
  algorithm to calculate convolution integrals in cosmological perturbation
  theory}},  {\em JCAP} {\bf 09} (2016) 015,
  [\href{http://arxiv.org/abs/1603.04826}{{\tt arXiv:1603.04826}}].

\bibitem{Simonovic:2017mhp}
M.~Simonovi\'c, T.~Baldauf, M.~Zaldarriaga, J.~J. Carrasco, and J.~A.
  Kollmeier, {\it {Cosmological perturbation theory using the FFTLog: formalism
  and connection to QFT loop integrals}},  {\em JCAP} {\bf 04} (2018) 030,
  [\href{http://arxiv.org/abs/1708.08130}{{\tt arXiv:1708.08130}}].

\bibitem{Fang:2019xat}
X.~Fang, E.~Krause, T.~Eifler, and N.~MacCrann, {\it {Beyond Limber: Efficient
  computation of angular power spectra for galaxy clustering and weak
  lensing}},  {\em JCAP} {\bf 05} (2020) 010,
  [\href{http://arxiv.org/abs/1911.11947}{{\tt arXiv:1911.11947}}].

\bibitem{2020arXiv200610256H}
C.~R. {Harris}, K.~{Jarrod Millman}, S.~J. {van der Walt}, R.~{Gommers},
  P.~{Virtanen}, D.~{Cournapeau}, E.~{Wieser}, J.~{Taylor}, S.~{Berg}, N.~J.
  {Smith}, R.~{Kern}, M.~{Picus}, S.~{Hoyer}, M.~H. {van Kerkwijk}, M.~{Brett},
  A.~{Haldane}, J.~{Fern{\'a}ndez del R{\'\i}o}, M.~{Wiebe}, P.~{Peterson},
  P.~{G{\'e}rard-Marchant}, K.~{Sheppard}, T.~{Reddy}, W.~{Weckesser},
  H.~{Abbasi}, C.~{Gohlke}, and T.~E. {Oliphant}, {\it {Array Programming with
  NumPy}},  {\em arXiv e-prints} (June, 2020) arXiv:2006.10256,
  [\href{http://arxiv.org/abs/2006.10256}{{\tt arXiv:2006.10256}}].

\bibitem{1972hmfw.book.....A}
M.~{Abramowitz} and I.~A. {Stegun}, {\em {Handbook of Mathematical Functions}}.
\newblock 1972.

\bibitem{mpmath}
F.~Johansson et~al., {\em mpmath: a {P}ython library for arbitrary-precision
  floating-point arithmetic (version 0.18)}, December, 2013.
\newblock {\tt http://mpmath.org/}.

\bibitem{Scipy:2020NatMe..17..261V}
P.~{Virtanen}, R.~{Gommers}, T.~E. {Oliphant}, M.~{Haberland}, T.~{Reddy},
  D.~{Cournapeau}, E.~{Burovski}, P.~{Peterson}, W.~{Weckesser}, J.~{Bright},
  S.~J. {van der Walt}, M.~{Brett}, J.~{Wilson}, K.~J. {Millman}, N.~{Mayorov},
  A.~R.~J. {Nelson}, E.~{Jones}, R.~{Kern}, E.~{Larson}, C.~J. {Carey},
  {\.I}.~{Polat}, Y.~{Feng}, E.~W. {Moore}, J.~{Vand erPlas}, D.~{Laxalde},
  J.~{Perktold}, R.~{Cimrman}, I.~{Henriksen}, E.~A. {Quintero}, C.~R.
  {Harris}, A.~M. {Archibald}, A.~H. {Ribeiro}, F.~{Pedregosa}, P.~{van
  Mulbregt}, and {SciPy 1. 0 Contributors}, {\it {SciPy 1.0: fundamental
  algorithms for scientific computing in Python}},  {\em Nature Methods} {\bf
  17} (Feb., 2020) 261--272, [\href{http://arxiv.org/abs/1907.10121}{{\tt
  arXiv:1907.10121}}].

\bibitem{Sympy:10.7717/peerj-cs.103}
A.~Meurer, C.~P. Smith, M.~Paprocki, O.~\v{C}ert\'{i}k, S.~B. Kirpichev,
  M.~Rocklin, A.~Kumar, S.~Ivanov, J.~K. Moore, S.~Singh, T.~Rathnayake,
  S.~Vig, B.~E. Granger, R.~P. Muller, F.~Bonazzi, H.~Gupta, S.~Vats,
  F.~Johansson, F.~Pedregosa, M.~J. Curry, A.~R. Terrel, v.~Rou\v{c}ka,
  A.~Saboo, I.~Fernando, S.~Kulal, R.~Cimrman, and A.~Scopatz, {\it Sympy:
  symbolic computing in python},  {\em PeerJ Computer Science} {\bf 3} (Jan.,
  2017) e103.

\bibitem{Villa:2015ppa}
E.~Villa and C.~Rampf, {\it {Relativistic perturbations in $\Lambda$CDM:
  Eulerian \& Lagrangian approaches}},  {\em JCAP} {\bf 1601} (2016), no.~01
  030, [\href{http://arxiv.org/abs/1505.04782}{{\tt arXiv:1505.04782}}].

\bibitem{Bernardeau:2001qr}
F.~Bernardeau, S.~Colombi, E.~Gaztanaga, and R.~Scoccimarro, {\it {Large scale
  structure of the universe and cosmological perturbation theory}},  {\em
  Phys.Rept.} {\bf 367} (2002) 1--248,
  [\href{http://arxiv.org/abs/astro-ph/0112551}{{\tt astro-ph/0112551}}].

\end{thebibliography}

\providecommand{\href}[2]{#2}\begingroup\raggedright\endgroup

\end{document}